\documentclass{aa}  

\usepackage{txfonts}

\usepackage{amsmath}
\usepackage{graphicx}
\usepackage{xcolor}
\usepackage{xparse}
\usepackage{bm}
\usepackage{latexsym}
\usepackage[
capitalize,
noabbrev
]{cleveref}
\usepackage{tikz}
\usepackage[normalem]{ulem}
\usepackage{paralist}
\usepackage{multirow}

\NewDocumentCommand{\curl}{o}{\ensuremath{\nabla \times #1}}
\NewDocumentCommand{\grad}{o o}{\ensuremath{\IfNoValueTF{#2}{\nabla}{\nabla_{#2}} #1}}
\NewDocumentCommand{\diverg}{o}{\ensuremath{\nabla \cdot #1}}

\NewDocumentCommand{\years}{o}{\ensuremath{
\IfNoValueF{#1}{#1 \,}
\mathrm{years}
}}

\NewDocumentCommand{\days}{o}{\ensuremath{
\IfNoValueF{#1}{#1 \,}
\mathrm{days}
}}

\NewDocumentCommand{\months}{o}{\ensuremath{
\IfNoValueF{#1}{#1 \,}
\mathrm{months}
}}

\NewDocumentCommand{\s}{o}{\ensuremath{
\IfNoValueF{#1}{#1 \,}
\mathrm{s}
}}

\NewDocumentCommand{\nT}{o}{\ensuremath{
\IfNoValueF{#1}{#1 \,}
\mathrm{nT}
}}

\NewDocumentCommand{\km}{o}{\ensuremath{
\IfNoValueF{#1}{#1 \,}
\mathrm{km}
}}

\NewDocumentCommand{\au}{o}{\ensuremath{
\IfNoValueF{#1}{#1 \,}
\mathrm{AU}
}}

\NewDocumentCommand{\amu}{o}{\ensuremath{
\IfNoValueF{#1}{#1 \,}
\mathrm{amu}
}}

\NewDocumentCommand{\temp}{o}{\ensuremath{
\IfNoValueF{#1}{#1 \times}
\mathrm{10^5 \; K}
}}

\NewDocumentCommand{\cc}{o}{\ensuremath{
\IfNoValueF{#1}{#1 \;}
\mathrm{cm}^{-3}
}}

\NewDocumentCommand{\pct}{o}{\ensuremath{
\IfNoValueF{#1}{#1 \;}
\%
}}

\NewDocumentCommand{\Rs}{o}{\ensuremath{
\IfNoValueF{#1}{#1 \;}
\mathrm{R_S}
}}

\NewDocumentCommand{\kms}{o}{\ensuremath{
\IfNoValueF{#1}{#1 \;}
\mathrm{km \, s^{-1}}
}}

\NewDocumentCommand{\mWcc}{o}{\ensuremath{
\IfNoValueF{#1}{#1 \;}
\mathrm{mW \cc}
}}

\NewDocumentCommand{\eV}{o}{\ensuremath{
\IfNoValueF{#1}{#1 \;}
\mathrm{eV}
}}

\NewDocumentCommand{\keV}{o}{\ensuremath{
\IfNoValueF{#1}{#1 \;}
\mathrm{keV}
}}

\NewDocumentCommand{\MeV}{o}{\ensuremath{
\IfNoValueF{#1}{#1 \;}
\mathrm{MeV}
}}

\NewDocumentCommand{\nucleon}{s o}{\ensuremath{
\IfNoValueF{#2}{#2 \;}
\IfBooleanTF{#1}{\mathrm{nucleon}}{\mathrm{nuc}}
}}

\NewDocumentCommand{\MeVnuc}{s o}{\ensuremath{
\IfNoValueF{#2}{#2 \;}
\MeV \! /\IfBooleanTF{#1}{\nucleon*}{\nucleon}
}}

\NewDocumentCommand{\keVe}{o}{\ensuremath{
\IfNoValueF{#1}{#1 \;}
\mathrm{keV/e}
}}

\NewDocumentCommand{\n}{o}{\ensuremath{n
\IfNoValueF{#1}{_{#1}}
}}

\NewDocumentCommand{\Element}{m}{\ensuremath{\mathrm{#1}}}
\NewDocumentCommand{\QState}{m m}{\ensuremath{\mathrm{#1}^{#2+}}}

\NewDocumentCommand{\Hy}{o}{\IfNoValueTF{#1}{\Element{H}}{\QState{H}{#1}}}
\NewDocumentCommand{\He}{o}{\IfNoValueTF{#1}{\Element{He}}{\QState{He}{#1}}}
\NewDocumentCommand{\C}{o}{\IfNoValueTF{#1}{\Element{C}}{\QState{C}{#1}}}
\NewDocumentCommand{\N}{o}{\IfNoValueTF{#1}{\Element{N}}{\QState{N}{#1}}}
\NewDocumentCommand{\Ox}{o}{\IfNoValueTF{#1}{\Element{O}}{\QState{O}{#1}}}
\NewDocumentCommand{\Ne}{o}{\IfNoValueTF{#1}{\Element{Ne}}{\QState{Ne}{#1}}}
\NewDocumentCommand{\Mg}{o}{\IfNoValueTF{#1}{\Element{Mg}}{\QState{Mg}{#1}}}
\NewDocumentCommand{\Si}{o}{\IfNoValueTF{#1}{\Element{Si}}{\QState{Si}{#1}}}
\NewDocumentCommand{\Su}{o}{\IfNoValueTF{#1}{\Element{S}}{\QState{S}{#1}}}
\NewDocumentCommand{\Ca}{o}{\IfNoValueTF{#1}{\Element{Ca}}{\QState{Ca}{#1}}}
\NewDocumentCommand{\Fe}{o}{\IfNoValueTF{#1}{\Element{Fe}}{\QState{Fe}{#1}}}

\NewDocumentCommand{\FIP}{o O{=}}{\ensuremath{
\ensuremath{\mathrm{FIP}}\IfNoValueF{#1}{\IfNoValueTF{#2}{=}{#2}\eV[11]}
}}

\NewDocumentCommand{\AbSEP}{O{X} O{\Ox}}{\ensuremath{#1/#2}}
\NewDocumentCommand{\AbSW}{s O{X} O{\Hy}}{\IfBooleanTF{#1}{\ensuremath{(#2/#3)}:\ensuremath{(#2/#3)_\mathrm{photo}}}
{\ensuremath{#2/#3}}}

\NewDocumentCommand{\PLawExp}{s o}{\ensuremath{b
\IfNoValueF{#2}{\IfBooleanTF{#1}{\approx}{=} #2}}}

\newcommand{\he}{\Element{He}}

\NewDocumentCommand{\M}{o}{\ensuremath{
\IfNoValueTF{#1}{\mathrm{M}}{(\mathrm{M})_{#1}}}}

\NewDocumentCommand{\Q}{o}{\ensuremath{
\IfNoValueTF{#1}{\mathrm{Q}}{(\mathrm{Q})_{#1}}}}

\NewDocumentCommand{\MpQ}{o}{\ensuremath{
\IfNoValueTF{#1}{\mathrm{M/Q}}{(\mathrm{M/Q})_{#1}}}}

\NewDocumentCommand{\As}{s o O{=} d__}{\ensuremath{\IfNoValueTF{#4}{A_s}{A_{s,#4}}
\IfNoValueF{#2}{#3 #2 \IfBooleanF{#1}{\%}}}}

\NewDocumentCommand{\vs}{s o O{=} d__}{\ensuremath{\IfNoValueTF{#4}{v_s}{v_{s,#4}}
\IfNoValueF{#2}{
\IfNoValueTF{#3}{=}{#3}
\IfBooleanTF{#1}{#2}{\kms[#2]}}
}}

\NewDocumentCommand{\vsw}{s o O{=}}{\ensuremath{v_\sw
\IfNoValueF{#2}{
\IfNoValueTF{#3}{=}{#3}
\IfBooleanTF{#1}{#2}{\kms[#2]}}
}}

\NewDocumentCommand{\vv}{s o O{=}}{\ensuremath{v_v
\IfNoValueF{#2}{
\IfNoValueTF{#3}{=}{#3}
\IfBooleanTF{#1}{#2}{\kms[#2]}}
}}

\NewDocumentCommand{\rc}{s o O{=}}{\ensuremath{r_c
\IfNoValueF{#2}{
\IfNoValueTF{#3}{=}{#3}
\IfBooleanTF{#1}{#2}{\au[#2]}}
}}

\NewDocumentCommand{\rA}{s o O{=}}{\ensuremath{r_A
\IfNoValueF{#2}{
\IfNoValueTF{#3}{=}{#3}
\IfBooleanTF{#1}{#2}{\au[#2]}}
}}

\NewDocumentCommand{\grate}{o o}{\ensuremath{
\gamma\IfNoValueF{#1}{/\Omega_{#1}}
\IfNoValueF{#2}{= 10^{{#2}}}
}}

\NewDocumentCommand{\gmax}{o}{\ensuremath{
\gamma_\mathrm{max}\IfNoValueF{#1}{/\Omega_{#1}}
}}

\NewDocumentCommand{\kvec}{o}{\ensuremath{
\vec{k} \rho\IfNoValueF{#1}{{_{#1}}}
}}

\NewDocumentCommand{\kpar}{o}{\ensuremath{
{k_\parallel} \rho\IfNoValueF{#1}{{_{#1}}}
}}

\NewDocumentCommand{\kper}{o}{\ensuremath{
{k_\perp} \rho\IfNoValueF{#1}{{_{#1}}}
}}

\NewDocumentCommand{\ani}{s o}{\ensuremath{
R\IfNoValueF{#2}{_{#2}}
\IfBooleanT{#1}{\, [\perp\!/\!\parallel]}
}}

\NewDocumentCommand{\Temp}{o}{\ensuremath{T{\IfNoValueF{#1}{_{#1}}}}}

\NewDocumentCommand{\Trat}{s m m o}{\ensuremath{
T_{\IfNoValueF{#4}{{#4};}#2}/T_{\IfNoValueF{#4}{{#4};}#3}
 \IfBooleanT{#1}{\, [\#]}
}}

\NewDocumentCommand{\pbeta}{s o}{\ensuremath{
\beta\IfNoValueF{#2}{_{#2}}
 \IfBooleanT{#1}{\, [\#]}
}}

\NewDocumentCommand{\pbetaR}{o}{\ensuremath{
(\pbeta[\parallel
\IfNoValueF{#1}{;#1}], \ani[#1])
}}

\NewDocumentCommand{\dv}{o}{\ensuremath{\Delta v\IfNoValueF{#1}{_{#1}}}}
\NewDocumentCommand{\ca}{o}{\ensuremath{C_{A\IfNoValueF{#1}{;#1}}}}
\NewDocumentCommand{\dvca}{o o}{\ensuremath{\dv[#1]/\ca[#2]}}

\NewDocumentCommand{\nuc}{o}{\ensuremath{\nu_{c\IfNoValueF{#1}{;#1}}}}
\NewDocumentCommand{\Nc}{o}{\ensuremath{N_{c\IfNoValueF{#1}{;#1}}}}
\NewDocumentCommand{\Ac}{o}{\ensuremath{A_{c\IfNoValueF{#1}{;#1}}}}

\NewDocumentCommand{\tauEXP}{o}{\ensuremath{
\tau_{\mathrm{exp}\IfNoValueF{#1}{;#1}
}}}

\NewDocumentCommand{\tauCC}{o}{\ensuremath{
\tau_{\mathrm{C}\IfNoValueF{#1}{;#1}
}}}

\NewDocumentCommand{\SSN}{o}{\ensuremath{\mathrm{SSN}
\IfNoValueF{#1}{#1}}}
\NewDocumentCommand{\NSSN}{o}{\ensuremath{\mathrm{NSSN}
\IfNoValueF{#1}{#1}}}

\newcommand{\sw}{\ensuremath{\mathrm{sw}}}

\NewDocumentCommand{\qpar}{o}{\ensuremath{
q_{\parallel
\IfNoValueF{#1}{;#1}
}}}

\NewDocumentCommand{\edv}{o}{\ensuremath{
\tilde{E}_{\dv[#1]
}}}

\NewDocumentCommand{\ndays}{o}{
\ensuremath{N_\mathrm{days}{\IfNoValueF{#1}{= {#1}}}}
}

\NewDocumentCommand{\se}{o}{\ensuremath{
S{\IfNoValueF{#1}{_{#1}}}
}}

\NewDocumentCommand{\ab}{o}{\ensuremath{
A{\IfNoValueF{#1}{_{#1}}}
}}

\NewDocumentCommand{\ahe}{o O{=}}{\ensuremath{\ab[\he]
\IfNoValueF{#1}{
\IfNoValueTF{#2}{=}{#2}#1\%}
}}

\NewDocumentCommand{\corr}{o}{\ensuremath{
\rho
\IfNoValueF{#1}{(#1)}
}}

\NewDocumentCommand{\xhel}{o O{=} o}{\ensuremath{\sigma_{c
\IfNoValueF{#3}{,#3}}
\IfNoValueF{#1}{
\IfNoValueTF{#2}{=}{#2}
#1}
}}

\NewDocumentCommand{\SpecInd}{o}{\ensuremath{\gamma
\IfNoValueF{#1}{_{#1}}}}
\NewDocumentCommand{\QT}{o}{\ensuremath{\mathrm{QT}
\IfNoValueF{#1}{= #1}}}

\NewDocumentCommand{\rsq}{o O{=}}{\ensuremath{R^2
\IfNoValueF{#1}{
\IfNoValueTF{#2}{=}{#2}
#1}
}}

\NewDocumentCommand{\rsqw}{o O{=}}{\ensuremath{R^2_w
\IfNoValueF{#1}{
\IfNoValueTF{#2}{=}{#2}
#1}
}}

\newcommand{\nst}[1]{#1\textsuperscript{st}}

\newcommand{\degree}{\ensuremath{^\circ}}

\definecolor{C0}{HTML}{1f77b4}
\definecolor{C1}{HTML}{ff7f0e}
\definecolor{C2}{HTML}{2ca02c}
\definecolor{C3}{HTML}{d62728}
\definecolor{C4}{HTML}{9467bd}
\definecolor{C5}{HTML}{8c564b}

\definecolor{QTFitGreen}{HTML}{2ca02c}
\definecolor{DodgerBlue}{HTML}{1e90ff}
\definecolor{Fuchsia}{HTML}{ff00ff}
\definecolor{TabGreen}{HTML}{2ca02c}
\definecolor{Cyan}{HTML}{00ffff}
\definecolor{LimeGreen}{HTML}{32cd32}
\definecolor{Lime}{HTML}{00ff00}

\definecolor{MaxPink}{HTML}{e377c2}
\definecolor{MinPurple}{HTML}{9467bd}

\definecolor{q}{HTML}{228B22}
\definecolor{wc}{HTML}{FF8C00}
\definecolor{dnc}{HTML}{FF00FF}
\definecolor{todo}{HTML}{e13748}
\definecolor{ben}{HTML}{e13748}
\definecolor{bob}{HTML}{228B22}

\NewDocumentCommand{\q}{s o m}{\IfBooleanF{#1}{\textcolor{q}{\textbf{Q}\IfNoValueF{#2}{ (#2)}: \textit{#3}}}}
\NewDocumentCommand{\answer}{s o m}{\IfBooleanF{#1}{\textcolor{q}{\textbf{A}\IfNoValueF{#2}{ (#2)}: \textit{#3}}}}

\NewDocumentCommand{\wc}{s m}{\IfBooleanTF{#1}{#2}{\textcolor{wc}{\textbf{WC:} \textit{#2}}}}
\NewDocumentCommand{\ws}{s m}{\IfBooleanTF{#1}{#2}{\textcolor{wc}{\textbf{WS:} \textit{#2}}}}

\NewDocumentCommand{\add}{s m}{\IfBooleanF{#1}{\textcolor{ben}{\textbf{Add:} \textit{#2}}}}
\NewDocumentCommand{\delete}{s m}{\IfBooleanF{#1}{\textcolor{todo}{\textbf{Delete:} \textit{#2}}}}

\NewDocumentCommand{\todo}{s o m}{\IfBooleanF{#1}{\textcolor{todo}{\textbf{TODO}\IfNoValueF{#2}{ (#2)}: \textit{#3}}}}
\NewDocumentCommand{\verify}{s o m}{\IfBooleanTF{#1}{#3}{\textcolor{todo}{\textbf{VERIFY}\IfNoValueF{#2}{ (#2)}: \textit{#3}}}}
\NewDocumentCommand{\goal}{s o m}{\IfBooleanTF{#1}{#3}{\textcolor{todo}{\textbf{GOAL}\IfNoValueF{#2}{ (#2)}: \textit{#3}}}}

\NewDocumentCommand{\move}{s o m}{\textcolor{dnc}{\textbf{\IfBooleanTF{#1}{Duplicate}{Move}}\IfNoValueF{#2}{ (#2)}: \textit{#3}}}
\NewDocumentCommand{\dupe}{o m}{\move*[#1]{#2}}
\NewDocumentCommand{\intro}{s m}{\IfBooleanTF{#1}{\dupe[Intro]{#2}}{\move[Intro]{#2}}}
\NewDocumentCommand{\dnc}{s m}{\IfBooleanTF{#1}{\dupe[DnC]{#2}}{\move[DnC]{#2}}}
\NewDocumentCommand{\fw}{s m}{\IfBooleanTF{#1}{\dupe[Future Work]{#2}}{\move[DnC]{#2}}}

\DeclareDocumentCommand{\EmptyTimes}{O{black}}{\ensuremath{\mathord{\begin{tikzpicture}[line width=0.2ex, x=1.5ex, y=1.5ex]
\draw[color=#1] (0, 0.25) -- (0.25, 0.5) -- (0, 0.75) -- (0.25, 1.0) -- (0.5, 0.75) -- (0.75, 1.0) -- (1.0, 0.75) -- (0.75, 0.5) -- (1.0, 0.25) -- (0.75, 0) -- (0.5, 0.25) -- (0.25, 0) -- cycle;
\end{tikzpicture}}}}

\DeclareDocumentCommand{\SolidBand}{O{black} D{<}{>}{1}}{\ensuremath{\mathord{\begin{tikzpicture}[line width=1.25ex, x=1.25ex, y=1.25ex, yshift=5ex]
\draw[color=#1, opacity=#2] (0,0.5) -- (1.5,0.5);
\draw[opacity=0, line width=0.1ex] (0,0) -- (1.5,0);
\end{tikzpicture}}}}

\DeclareDocumentCommand{\SolidBandVertLines}{O{black} D{<}{>}{1}}{\ensuremath{\mathord{\begin{tikzpicture}[line width=1.25ex, x=1.25ex, y=1.25ex, yshift=5ex]
\draw[color=#1, opacity=#2] (0,0.5) -- (1.5,0.5);
\draw[opacity=0, line width=0.1ex] (0,0) -- (1.5,0);
\draw[color=#1, line width=0.2ex] (0, 0) -- (0, 1);
\draw[color=#1, line width=0.2ex] (1.5, 0) -- (1.5, 1);
\end{tikzpicture}}}}

\DeclareDocumentCommand{\SolidLine}{O{black}}{\ensuremath{\mathord{\begin{tikzpicture}[line width=0.3ex, x=1.25ex, y=1.25ex, yshift=5ex]
\draw[color=#1] (0,0.5) -- (1,0.5);
\draw[opacity=0] (0,0) -- (1,0);
\end{tikzpicture}}}}

\DeclareDocumentCommand{\DashedLine}{O{black} o D{<}{>}{1}}{\ensuremath{\mathord{\begin{tikzpicture}[line width=0.3ex, x=1.25ex, y=1.25ex, yshift=5ex]
\IfNoValueF{#2}{\draw[color=#2, opacity=#3] (0, 0.5) -- (1.75, 0.5);}
\draw[color=#1, opacity=#3] (0,0.5) -- (0.75,0.5);
\draw[color=#1, opacity=#3] (1.0,0.5) -- (1.75,0.5);
\draw[opacity=0] (0,0) -- (1.25,0);
\end{tikzpicture}}}}

\DeclareDocumentCommand{\HighlightDashedLine}{O{black} O{LimeGreen} D{<}{>}{1}}{\ensuremath{\mathord{\begin{tikzpicture}[line width=0.3ex, x=1.25ex, y=1.25ex, yshift=5ex]
\draw[color=#2, opacity=#3, line width=0.8ex] (0, 0.5) -- (1.75, 0.5);
\draw[color=#1] (0,0.5) -- (0.75,0.5);
\draw[color=#1] (1.0,0.5) -- (1.75,0.5);
\draw[opacity=0] (0,0) -- (1.25,0);
\end{tikzpicture}}}}

\DeclareDocumentCommand{\DotDotDotLine}{O{white} O{black}}{\ensuremath{\mathord{\begin{tikzpicture}[line width=0.3ex, x=1.25ex, y=1.25ex, yshift=5ex]
\draw[color=#1] (0, 0.5) -- (1.5, 0.5);
\draw[color=#2] (0,0.5) -- (0.3,0.5);
\draw [color=#2](0.6,0.5) -- (0.9,0.5);
\draw[color=#2] (1.2,0.5) -- (1.5,0.5);
\draw[opacity=0] (0,0) -- (1.5,0);
\end{tikzpicture}}}}

\DeclareDocumentCommand{\DotDotDotLineTwo}{O{yellow} O{black}}{\ensuremath{\mathord{\begin{tikzpicture}[line width=0.3ex, x=1.25ex, y=1.25ex, yshift=5ex]
\draw[color=#1] (0,0.5) -- (1.5,0.5);
\draw[color=#2] (0,0.5) -- (0.3,0.5);
\draw[color=#2] (0.6,0.5) -- (0.9,0.5);
\draw[color=#2] (1.2,0.5) -- (1.5,0.5);
\draw[opacity=0] (0,0) -- (1.5,0);
\end{tikzpicture}}}}

\DeclareDocumentCommand{\DashDotDotDotLine}{O{white} O{black}}{\ensuremath{\mathord{\begin{tikzpicture}[line width=0.3ex, x=1.25ex, y=1.25ex, yshift=5ex]
\draw[color=#1] (0, 0.5) -- (2.8, 0.5);
\draw[color=#2] (0,0.5) -- (1,0.5);
\draw[color=#2] (1.3,0.5) -- (1.6,0.5);
\draw[color=#2] (1.9,0.5) -- (2.2,0.5);
\draw[color=#2] (2.5,0.5) -- (2.8,0.5);
\draw[opacity=0] (0,0) -- (2.8,0);
\end{tikzpicture}}}}

\DeclareDocumentCommand{\Circle}{O{black}}{\ensuremath{\mathord{\begin{tikzpicture}[line width=0.3ex, x=1.25ex, y=1.25ex, yshift=5ex]
\draw[color=#1] circle (0.75ex);
\end{tikzpicture}}}}

\NewDocumentCommand{\sect}{o m}{Section~\ref{sec:#2}\IfNoValueF{#1}{ #1}}
\NewDocumentCommand{\eq}{o m}{\cref{eq:#2}\IfNoValueF{#1}{ #1}}
\NewDocumentCommand{\tbl}{o m}{\cref{tbl:#2}\IfNoValueF{#1}{ #1}}

  \usepackage{multirow}
\usepackage{tablefootnote}

\NewDocumentCommand{\plotVswAb}{s}{
\IfBooleanTF{#1}{\begin{figure*}}{\begin{figure}}
\begin{centering}
\includegraphics[page=10,width=\linewidth]{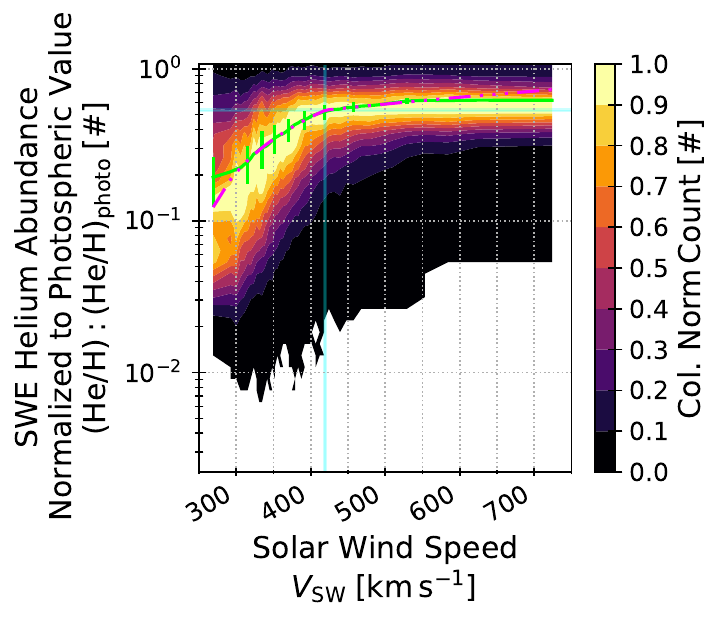}
\caption{\label{fig:VswAb}
A contour plot corresponding to a column-normalized 2D histogram of the SWE helium abundance as a function of the proton speed observed at \emph{Wind}.
The solid green line and error bars are the mean and standard deviation in each column.
The pink dash-dotted line show the result of bi-linear fit to the green line, where each line is selected as the minimum of both lines in the bi-linear function over the full domain.
Only speed \vsw[300][>] are included in the fit.
Semi-transparent blue lines indicate the saturation speed (\vs) and saturation abundance (\As) along with their uncertainties, where the bi-linear function changes slope.
}
\end{centering}
\IfBooleanTF{#1}{\end{figure*}}{\end{figure}}
}

\NewDocumentCommand{\plotVswAbComparison}{s}{
\IfBooleanTF{#1}{\begin{figure*}}{\begin{figure}}
\begin{centering}
\includegraphics[width=\IfBooleanTF{#1}{0.7\linewidth}{\linewidth}]{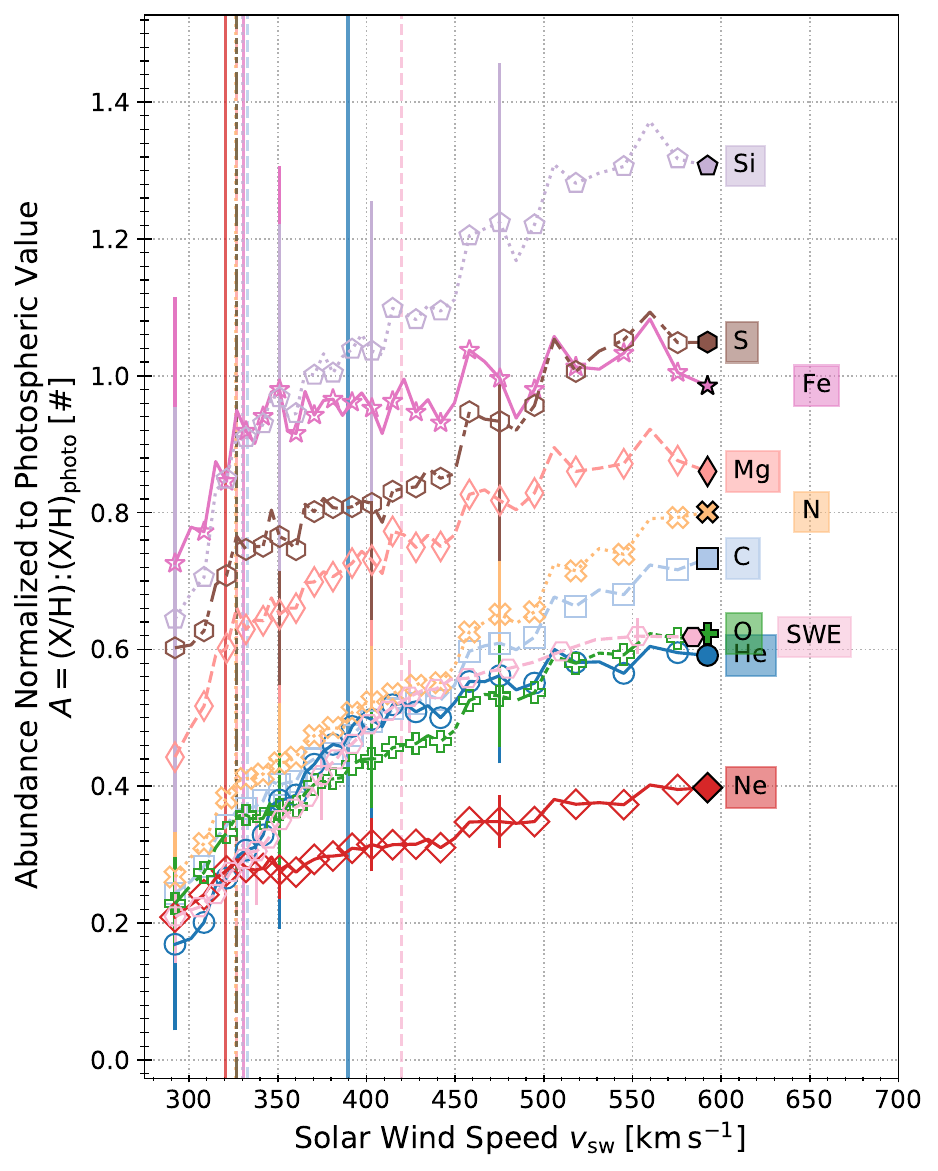}
\caption{\label{fig:VswAbComparison}
Abundances averaged in solar wind speed bins.
Saturation speeds (\vs) are indicated by vertical lines of the corresponding color.
The species are indicated on the right hand side of the plot.
The SWE observations of \ahe\ from \cref{fig:VswAb} are shown for reference in pink hexagons and labeled \textit{SWE}.
Only every second bin is marked for visual clarity.
}
\end{centering}
\IfBooleanTF{#1}{\end{figure*}}{\end{figure}}
}

\NewDocumentCommand{\plotVswAbComparisonScaled}{s}{
\IfBooleanTF{#1}{\begin{figure*}}{\begin{figure}}
\begin{centering}
\includegraphics[width=\IfBooleanTF{#1}{0.7\linewidth}{\linewidth}]{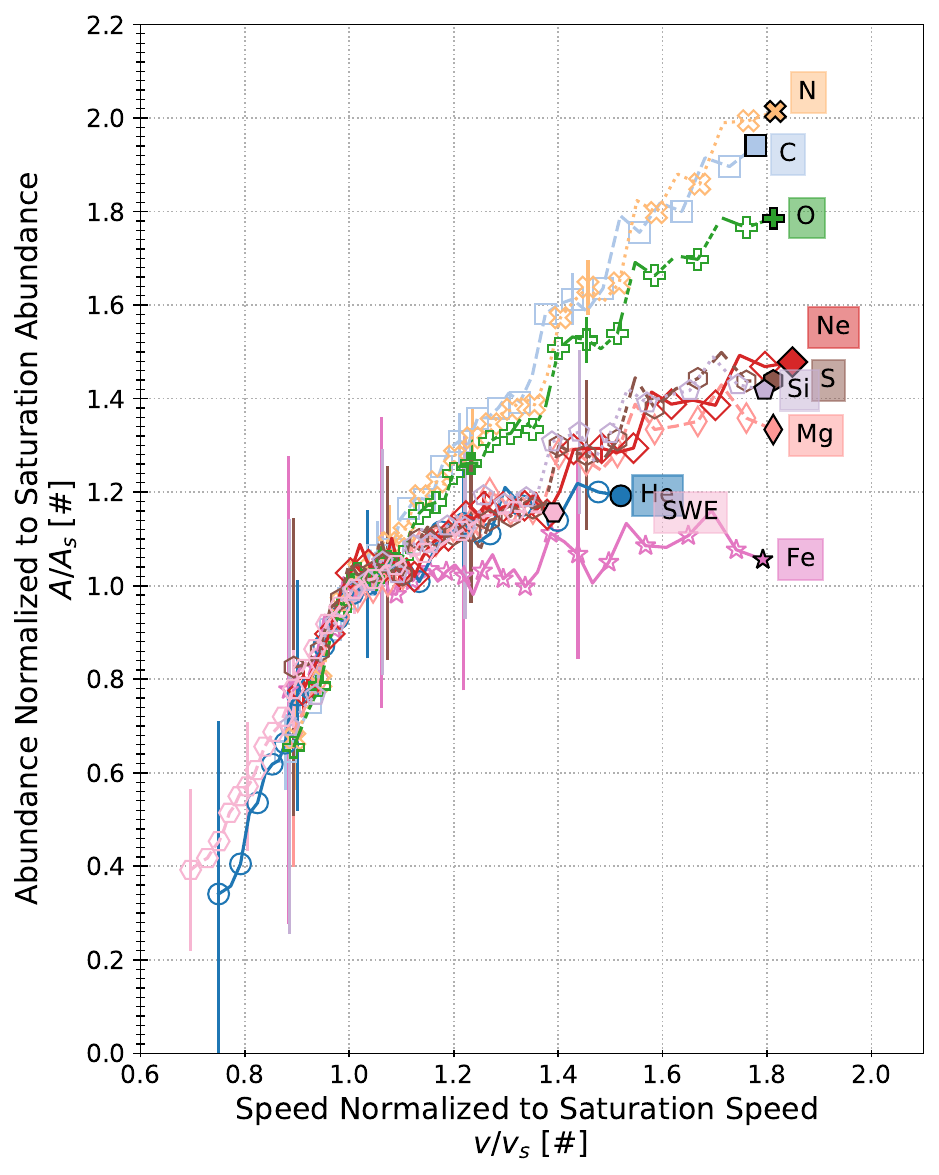}
\caption{\label{fig:VswAbComparison:scaled}
Observations plotted in \cref{fig:VswAbComparison}, but scaled to their $(\vs, \As)$ values, which is plotted at $(1,1)$.}
\end{centering}
\IfBooleanTF{#1}{\end{figure*}}{\end{figure}}
}

\NewDocumentCommand{\plotHingeVswComparison}{s}{
\IfBooleanTF{#1}{\begin{figure*}}{\begin{figure}}
\begin{centering}
\includegraphics[width=\linewidth]{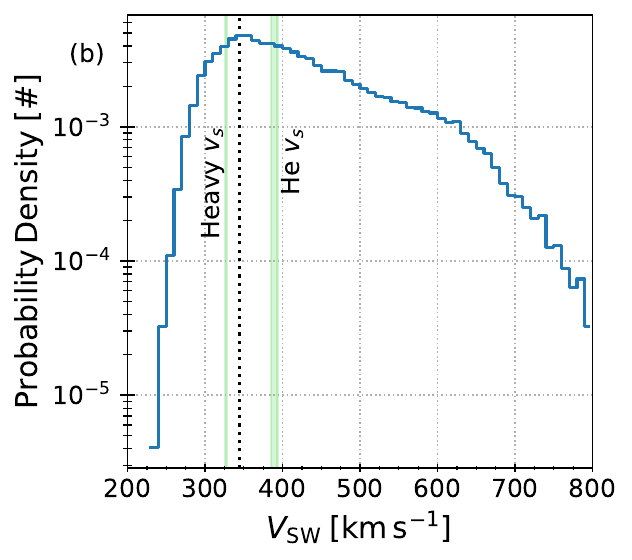}
\caption{\label{fig:xh-vsw-comparison}
Probability density of \vsw\ observed by SWICS and SWE.
The vertical green lines are saturation speeds \vs\ including uncertainty.
The vertical black dotted line is the peek solar wind speed bin, \vsw[345].}
\end{centering}
\IfBooleanTF{#1}{\end{figure*}}{\end{figure}}
}

\NewDocumentCommand{\PlotVsChemistry}{s}{
\IfBooleanTF{#1}{\begin{figure*}}{\begin{figure}}
\begin{centering}
\includegraphics[page=41, width=\linewidth]{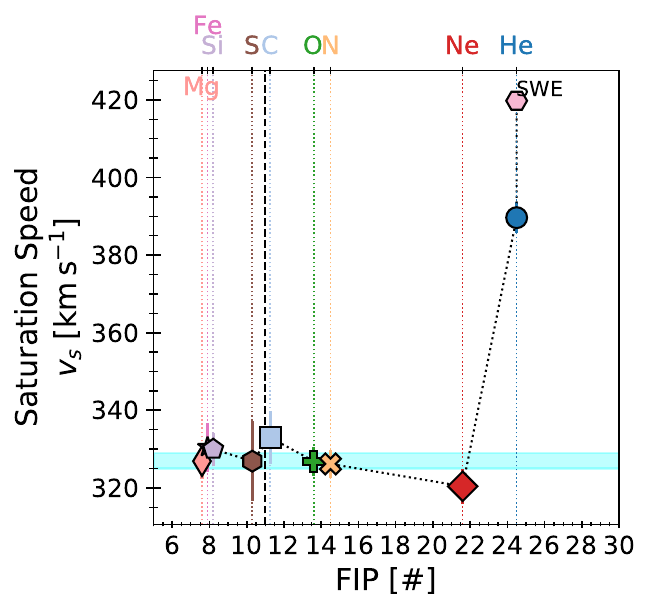}
\caption{\label{fig:vs-chemistry}
The saturation speed (\vs) as a function of first ionization potential (\FIP).
The vertical dashed line is \eV[11], the nominal change between high and low \FIP.
The semi-transparent, horizontal blue bar indicates the weighted average of \vs[327 \pm 2] for elements heavier than \He.
To within their mutual uncertainties, all but \Ne\ and \Si\ have the same \vs.
}
\end{centering}
\IfBooleanTF{#1}{\end{figure*}}{\end{figure}}
}

\NewDocumentCommand{\PlotAsChemistry}{s}{
\IfBooleanTF{#1}{\begin{figure*}}{\begin{figure}}
\begin{centering}
\includegraphics[page=23, width=\linewidth]{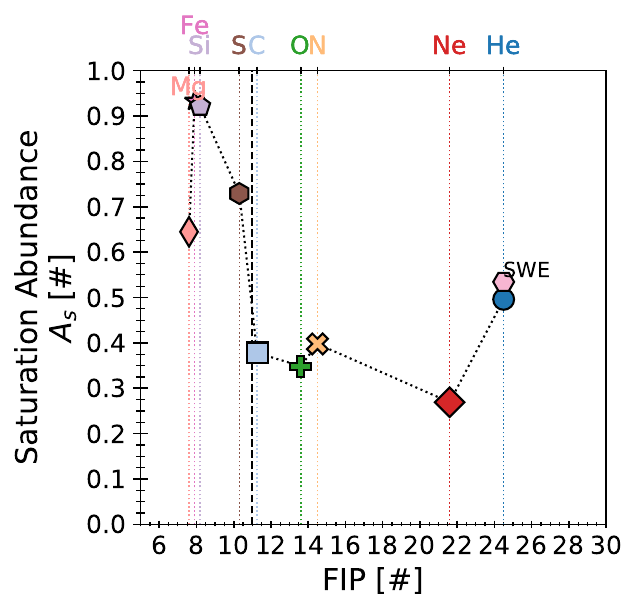}
\caption{\label{fig:as-chemistry}
Each element's saturation abundance (\As) as a function of FIP.
The vertical dashed line is \eV[11], the nominal change between high and low \FIP.
The $\sim2\times$ difference between low and high \FIP\ abundances is expected.
}
\end{centering}
\IfBooleanTF{#1}{\end{figure*}}{\end{figure}}
}

\NewDocumentCommand{\PlotVvChemistry}{s}{
\IfBooleanTF{#1}{\begin{figure*}}{\begin{figure}}
\begin{centering}
\includegraphics[page=33, width=\linewidth]{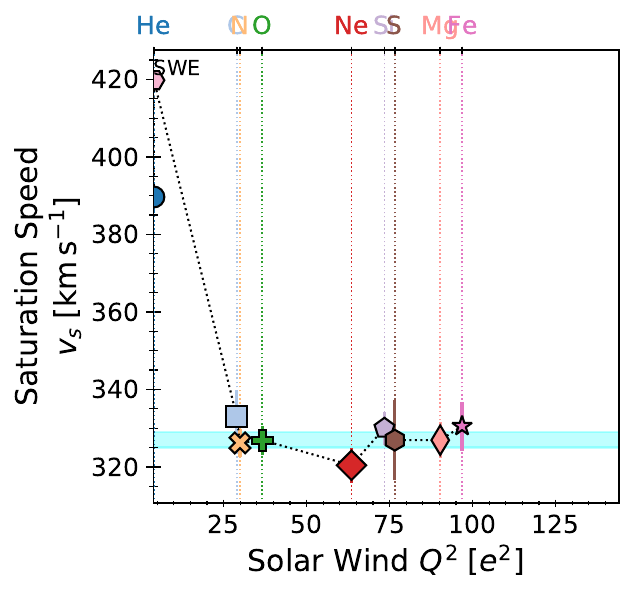}
\caption{\label{fig:vv-chemistry}
The vanishing speed (\vv) as a function of solar wind charge state.
The blue band is the solar wind's vanishing speed from \citet{Kasper2006}.
\dnc{Verify math: If we decrease the vanishing speed by \kms[63 \pm 5], the shift in \vs\ between heavy and \He\ \vs, then the speeds coincide. Further substantiating an overall dynamic relevance for \He\ in solar wind reaching its asymptotic speed.}
}
\end{centering}
\IfBooleanTF{#1}{\end{figure*}}{\end{figure}}
}

\NewDocumentCommand{\PlotFastestMass}{s}{
\IfBooleanTF{#1}{\begin{figure*}}{\begin{figure}}
\begin{centering}
\includegraphics[page=2, width=\linewidth]{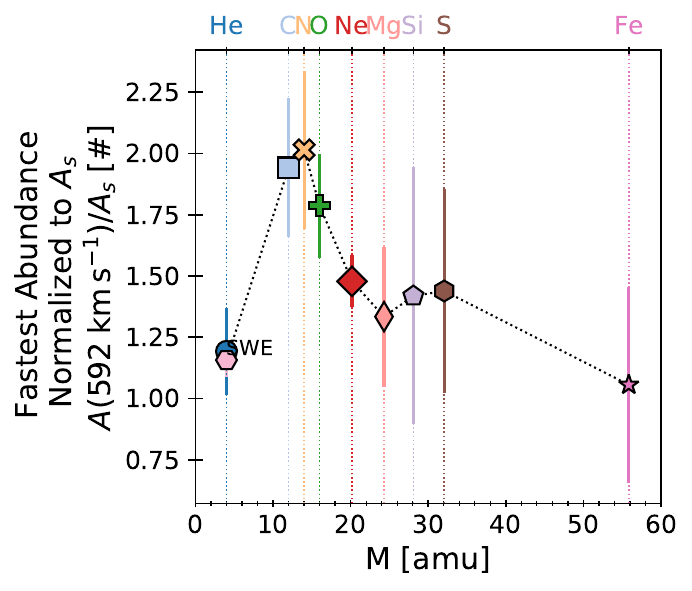}
\caption{\label{fig:af-chemistry}
The abundance at \vsw[592] normalized to \As\ as a function of element mass.
Excluding \He, the decreasing trend with increasing \M\ indicates a heavy ion fractionation process in fast solar wind.
}
\end{centering}
\IfBooleanTF{#1}{\end{figure*}}{\end{figure}}
}

\NewDocumentCommand{\PlotFastestQState}{s}{
\IfBooleanTF{#1}{\begin{figure*}}{\begin{figure}}
\begin{centering}
\includegraphics[page=13, width=\linewidth]{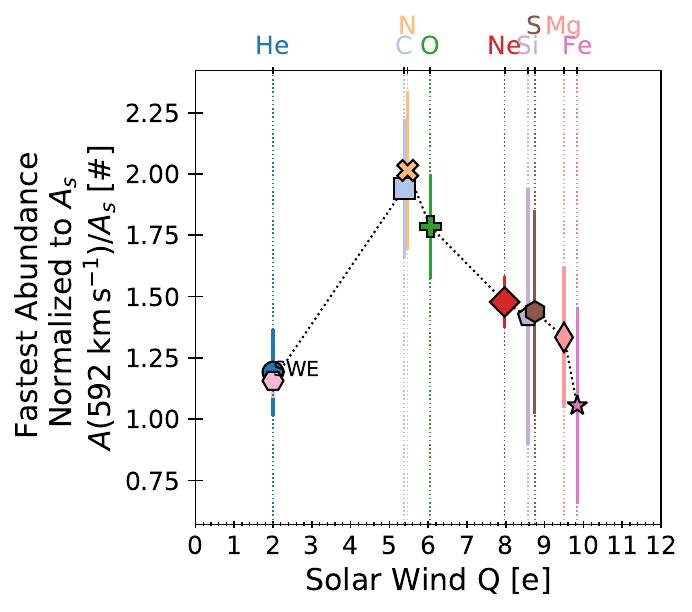}
\caption{\label{fig:af-qstate}
The abundance at \vsw[592] normalized to \As\ as a function of solar wind charge state.
Excluding \He, the decreasing trend with increasing \M\ indicates a heavy ion fractionation process in fast solar wind.
}
\end{centering}
\IfBooleanTF{#1}{\end{figure*}}{\end{figure}}
}

\newcommand{\eqBilinear}{
\begin{equation} \label{eq:bilinear}
A(v) = \min\left[A_1(v),A_2(v)\right] = \min\left[m_1(v - v_1), m_2 (v - v_2)\right].
\end{equation}
}

\newcommand{\eqSspeed}{
\begin{equation} \label{eq:saturation}
v_\mathrm{s} = \frac{m_2 v_2 - m_1 v_1}{m_2 - m_1}.
\end{equation}
}

\clearpage{}

\clearpage{}
\clearpage{}

\newcommand{\TblSaturation}{
\begin{table*}
\centering
\begin{tabular}{c c c c c c c}
\hline
\hline
{} & Vanishing & Saturation & Saturation & Fast Wind & Fastest & Slow Wind \\
{} & Speed & Speed & Abundance & Slope & Abundance & Observations \\
{} & $\vv$& \vs & \As & $m_i$ & $A\left(592 \, \mathrm{km \, s^{-1}}\right)$ & $v < \vs$ \\
{} & $[\kms]$ & $[\kms]$ & $[\#]$ & $[\mathrm{\% \, km^{-1} s}]$ & $[\#]$ & $[\%]$ \\
\hline
SWE & $237 \pm 6$ & $399 \pm 2$ & $0.520 \pm 0.004$ & $0.00045 \pm 0.00003$ & $0.60 \pm 0.04$ & 50 \\
He & $243 \pm 3$ & $390 \pm 4$ & $0.496 \pm 0.009$ & $0.00055 \pm 0.00007$ & $0.59 \pm 0.09$ & 44 \\
C & $220 \pm 7$ & $333 \pm 7$ & $0.378 \pm 0.013$ & $0.00152 \pm 0.00005$ & $0.73 \pm 0.10$ & 18 \\
N & $224 \pm 6$ & $326 \pm 4$ & $0.398 \pm 0.008$ & $0.00158 \pm 0.00004$ & $0.80 \pm 0.13$ & 17 \\
O & $228 \pm 6$ & $327 \pm 4$ & $0.349 \pm 0.008$ & $0.00115 \pm 0.00004$ & $0.62 \pm 0.07$ & 18 \\
Ne & $196 \pm 17$ & $320 \pm 4$ & $0.269 \pm 0.004$ & $0.00048 \pm 0.00002$ & $0.40 \pm 0.03$ & 18 \\
Mg & $221 \pm 13$ & $327 \pm 5$ & $0.645 \pm 0.011$ & $0.00108 \pm 0.00007$ & $0.86 \pm 0.18$ & 17 \\
Si & $214 \pm 12$ & $330 \pm 4$ & $0.921 \pm 0.015$ & $0.00178 \pm 0.00008$ & $1.31 \pm 0.48$ & 17 \\
S & $156 \pm 45$ & $327 \pm 10$ & $0.729 \pm 0.018$ & $0.00132 \pm 0.00008$ & $1.05 \pm 0.30$ & 14 \\
Fe & $168 \pm 34$ & $331 \pm 6$ & $0.933 \pm 0.012$ & $0.00034 \pm 0.00009$ & $0.99 \pm 0.37$ & 16 \\
\hline
Avg & $221 \pm 3$ & $327 \pm 2$ & $0.383 \pm 0.003$ & $0.00083 \pm 0.00001$ & --- & --- \\
Low FIP & $212 \pm 8$ & $329 \pm 3$ & $0.795 \pm 0.007$ & $0.00115 \pm 0.00004$ & --- & --- \\
High FIP & $223 \pm 4$ & $326 \pm 2$ & $0.302 \pm 0.003$ & $0.00079 \pm 0.00002$ & --- & --- \\
\hline
\hline
\end{tabular}
\caption{\label{tbl:saturation}
Saturation speeds and abundances along with the slow wind x-intercept (\vv, vanishing speed) and fast wind slope ($m_i$).
These parameters characterize the fits to \cref{eq:bilinear}.
The abundance observed at \kms[592] is given by $A\left(\kms[592]\right)$.
All abundances are normalized to their photospheric value.
SWE is \AbSW[\He] observed at \emph{Wind}.
The average value is calculated excluding SWE and SWICS \AbSW[\He].
High and Low FIP averages exclude \AbSW[\He] as well.
Percentage of data in slow wind regions ($v < \vs$) shows that non-trivial portions of the observations occur at these speeds.
}
\end{table*}
}

\clearpage{}
\clearpage{}

\NewDocumentCommand{\TblDelay}{s}{
\IfBooleanTF{#1}{\begin{table*}}{\begin{table}}
\centering
\begin{tabular}{ccccccccc}
\hline
{} & \multicolumn{4}{c}{Slow Wind} & \multicolumn{4}{c}{Fast Wind} \\
{} &  SWE & Observed & Peak & Delay &  SWE & Observed & Peak & Delay \\
{} & [\#] & [\#] & [\#] & [Days] & [\#] & [\#] & [\#] & [Days] \\
\hline
SWE &  --- &     0.92 & 0.92 &    13 &  --- &     0.76 & 0.80 &   233 \\
He  & 0.83 &     0.81 & 0.84 &   -95 & 0.54 &     0.49 & 0.50 &   -14 \\
C   & 0.47 &     0.49 & 0.61 &  -292 &  --- &      --- &  --- &   --- \\
N   & 0.66 &     0.62 & 0.70 &  -209 &  --- &      --- &  --- &   --- \\
O   & 0.78 &     0.72 & 0.75 &  -170 &  --- &      --- &  --- &   --- \\
Ne  & 0.85 &     0.71 & 0.74 &  -173 &  --- &      --- &  --- &   --- \\
Mg  & 0.87 &     0.75 & 0.77 &  -172 & 0.46 &     0.53 & 0.57 &  -256 \\
Si  & 0.79 &     0.77 & 0.82 &  -176 & 0.52 &     0.51 & 0.53 &   -54 \\
S   & 0.86 &     0.76 & 0.78 &  -169 &  --- &      --- &  --- &   --- \\
Fe  & 0.91 &     0.81 & 0.83 &   -51 & 0.58 &     0.59 & 0.63 &   -91 \\
\hline
\end{tabular}
\caption{\label{tbl:delay}
Fast and slow wind correlation coefficients and relevant statistics for the species identified in the first column.
The columns labeled \emph{SWE} provide the correlation coefficient between a given species and \ahe\ observed by the SWE Faraday cups.
Only significant coefficients with p-values $< 0.05$ are shown.
Other columns provide statistics on the observed and peak correlation coefficients between \AbSW*\ and \SSN.
The delay is the days by which \SSN\ is shifted to derive the peak coefficient.
\question{Do we really want to show the delay? We are limited in the time period of observations, so I don't trust the delays as much.}
\question{Do we want to only show slow wind? The correlations between SWE \AbSW[\He] and SWICS abundances could justify that the fast wind is sufficiently steady that we just don't need to report this.}
}
\IfBooleanTF{#1}{\end{table*}}{\end{table}}
}
\clearpage{}

\graphicspath{{./}{figures/}}
\DeclareGraphicsExtensions{.pdf,.png,.jpg,.eps,}

\newcommand{\heavy}{\ensuremath{\mathrm{Heavy}}}

\usepackage{lipsum}
\NewDocumentCommand{\BlindText}{O{6}}{\todo{Remove blind text. This is here to help figures render nicely.}
\textcolor{white}{\lipsum*[1-#1]}}

\begin{document}

\newcommand{\satpoint}{\ensuremath{\left(\vs,\As\right)}}

\title{On the transition from Slow to Fast Wind as Observed in Composition Observations}

\newcommand{\SwRI}{Space Science and Engineering \\
Southwest Research Institute \\
6220 Culebra Road \\
San Antonio, TX 78238, USA
}

\newcommand{\CfA}{Center for Astrophysics $\vert$ Harvard \& Smithsonian,
\\60 Garden Street, Cambridge, MA 02138, USA}

\newcommand{\CLaSP}{University of Michigan \\
Department of Climate and Space Sciences \& Engineering \\
Climate and Space Research Building \\
2455 Hayward St.\\
Ann Arbor, MI, 48109, USA}

\author{B.\ L.\ Alterman\inst{1}
\and
Y.\ J.\ Rivera\inst{2}
\and
S.\ T.\ Lepri\inst{3}
\and
J.\ M.\ Raines \inst{3}
}

\institute{\SwRI\\
\email{blalterman@swri.org}
\and
\CfA
\and
\CLaSP
}

\date{Received September 15, 1996; accepted March 16, 1997}

  \abstract{The solar wind is typically categorized as fast and slow based on the measured speed (\vsw).
The separation between these two regimes is often set between 400 and \kms[600] without a rigorous definition.
Observations with \vsw\ above this threshold are considered ``fast'' and are typically considered to come from polar regions, i.e.~coronal holes.
Observations with \vsw\ below this threshold speed are considered ``slow'' wind and typically considered to originate outside of solar coronal holes.
Observations of the solar wind's kinetic signatures, chemical makeup, charge state properties, and Alfvénicity suggest that such a two-state model may be insufficiently nuanced to capture the relationship between the solar wind and its solar sources.
As heavy ion composition ratios are unchanged once the solar wind leaves the Sun, they serve as a key tool for connecting \emph{in situ} observations to their solar sources.
Helium (\He) is the most abundance solar wind ion heavier than hydrogen (\Hy).
Long duration observations from the Wind Solar Wind Experiment (SWE) Faraday cups show that the solar wind helium abundance has two distinct gradients at speeds above and below \kms[\sim400].
This is a key motivator for identifying the separation between fast and slow wind at such a speed.
}{
We test this two-state fast/slow solar wind paradigm with heavy ion abundances (\AbSW) and characterize how the transition between fast and slow wind states impacts heavy ion in the solar wind.
}{
We study the variation of the gradients of the helium and heavy ion abundances as a function of solar wind speed and characterize how the gradient of each abundance changes in fast and slow wind.
We calculate \vsw\ as the proton or hydrogen bulk speed.
The work uses \emph{Advanced Composition Explorer} (ACE) heavy ion observations collected by the \emph{Solar Wind Ion Composition Spectrometer} (SWICS) from 1998 to 2011.
We compare the helium abundance observed by ACE/SWICS to the helium abundance observed by Wind/SWE to show the results are consistent with prior work.
}{
We show that (1) the speed at which heavy ion abundances indicate a change between fast and slow solar wind as a function of speed is slower than the speed indicated by the helium abundance; (2) this speed is independent of heavy ion mass and charge state; (3) the abundance at which heavy ions indicate the transition between fast and slow wind is consistent with prior observations of fast wind abundances; (4) and there may be a mass or charge-state dependent fractionation process present in fast wind heavy ion abundances.}{
We infer that (1) identifying slow solar wind as having a speed \vsw[400][\lesssim] may mix solar wind from polar and equatorial sources; (2) \He\ may be impacted by the acceleration necessary for the solar wind to reach the asymptotic fast, non-transient values observed at \au[1]; and (3) heavy ions are fractionated in the fast wind by a yet-to-be-determined mechanism.
}


   \maketitle

\section{Introduction \label{sec:intro}}

The solar wind is a magnetized plasma originating in the solar atmosphere, where
it is both energized and accelerated by the Sun before continuously flowing into and permeating interplanetary space.
The class of surface feature from which a given solar wind stream emerges determines the acceleration and energization mechanisms and therefore profiles \citep{Viall:9Q}.
The variation in solar wind ``types'' or classes was first established from its bi-modal speed profile, observed during solar minima from spacecraft with orbits in the ecliptic plane.
However, Ulysses firmly established the difference between ``fast'' and ``slow'' wind when the spacecraft's passages over the Sun's polar regions revealed a markedly higher speed (\vsw[400-500][>] at latitude $> 35\degree$) compared to lower latitudes near the streamer belt \citep{McComas2008, vonSteiger2000}.
These results demonstrated that the higher speed wind observed--even in the ecliptic--originates from structures with continuously open magnetic field structures on the Sun (i.e.~deep within coronal holes), while the slower speed wind arose from closed field structures that are intermittently open to the heliosphere, such as active regions, helmet streamers, the Quiet Sun, and pseudostreamers \citep{Fisk1999,Subramanian2010,Antiochos2011,Crooker2012,Abbo2016,Antonucci2005,DelZanna2019,Doschek2019}.
The frequency at which various source regions occur and the regions on the Sun where they commonly occur varies with solar activity \citep{McIntosh2015a,Hewins2020,Wang2002,Tlatov2014,Hathaway2015}.
These variations with solar activity impact in situ observations at \au[1], especially in regards to the occurrence rate of slow and fast wind along with the other features typically associated with them \citep{Hirshberg1973,Alterman2019,Alterman2021,Yogesh2023,McComas2008,Marsch2006b,DAmicis2015,Zerbo2015,Schwenn2006,Nicolaou2014,Du2012,ACE:SWICS:AUX}.
However, the contribution of individual source regions to the slow solar wind and how these contributions change with solar cycle is still a major open question in heliophysics.

The solar wind becomes supersonic near the Sun where thermal energy is converted to kinetic energy \citep{Parker1958a,Meyer-Vernet2007a}.
Above this height, it is further accelerated to an asymptotically faster speed during propagation through interplanetary space \citep{Leer1980,Hansteen2012,Holzer1981,Holzer1980a,Johnstone2015}.
Broadly, the solar wind's asymptotic speed is a distinguishing characteristic of the variation of solar source regions.
However within a small speed range, the solar wind can vary in density, temperature, Alfvénicity, chemical makeup, charge state population, and kinetic signatures, providing additional insight to its coronal origin and early development \citep{vonSteiger2000,Geiss1995,Geiss1995b,Zhao2022,Xu2014,Fu2017,Fu2015}.
The charge state and elemental composition of the solar wind are directly related to the temperature and density profiles of the solar source regions. 
Above a few solar radii, the charge state and elemental abundances remain fixed.
As such, these two properties are distinguishing tracers of its solar origin \citep{Xu2014,Zhao:InSituComposition:Sources,Zhao2017a}, which includes identifying the boundaries between solar wind streams of different origins as well as a means of identifying boundaries of transients \citep{Zurbuchen2006,Richardson2010}.
In other words, heavy ion observations provide a direct connection between \emph{in situ} observations and the solar wind's properties at its solar origin.
\citet{Rivera2022a} summarize the details of heavy ion properties at the Sun and in the solar wind.

Properties of elemental composition in the corona are observed to vary among neighboring coronal structures and often differ from the composition of the Sun's photosphere \citep{Pottasch1963,Feldman2000,Widing2001,Brooks2015}.
A main elemental fractionation process is thought to be driven by the reflection and refraction of Alfvén waves at the chromosphere-corona boundary \citep{LR:FIP}.
The resulting outward directed pondermotive force preferentially transport charged particles from the chromosphere to the corona while neutrals that are not yet ionized unaffected.
Within the associated fractionation timescale, elements with a low first ionization potential (FIP $< 11$eV) are appreciably enhanced in the corona while those with high FIP ($> 11$eV) remain at photospheric levels.
This is referred to as the ``FIP effect''.
Because the behavior of this pondermotive force and the magnitude of its impact on elemental composition varies with magnetic topology at the Sun, elemental abundances measured at the Sun and heliosphere are directly linked to the fractionation phenomena across different source regions.
In other words, elemental composition can indicate magnetic topology and field strength, thermal structure, and loop confinement duration, the latter of which is related to gravitational stratification \citep{Raymond1997,Feldman2000,Widing2001,Laming2004,Laming2012,Weberg2015,Rivera2022,Baker2023,Mihailescu2023}.

Ions heavier than \He\ measured in the heliosphere exhibit strong FIP effect fractionation with slower speed wind being composed of a range of low-FIP enhanced plasma (by factors of greater than 3) while fast speed wind converges to abundances more similar to the Sun's photospheric composition \citep{vonSteiger2000}.
The "FIP bias" is the ratio of high FIP abundances to low FIP abundances.
Polar passes of Ulysses observations find that the fast solar wind has a steady ion and elemental composition while the slow solar wind can reflect fast wind characteristics as well \citep{Stakhiv2015}. 
The slow wind was sub-characterized as typical slow wind and boundary wind where the boundary wind contained ionic composition similar to slow wind but elemental composition resembling fast wind.
The heavy ion variability of the slow wind is also observed on the ecliptic plane with ACE/SWICS observations \citep{Livi2003}. Similarly, the ion composition ($\Ox[7]/\Ox[6]$, $\C[6]/\C[5]$, $\C[6]/\C[5]$) have been used as tracers of solar wind origin back the Sun \citep{Zhao2017a}.
When organized by bulk speed, the $\Ox[7]/\Ox[6]$ ratio has a large overlap between traditionally fast (coronal hole) and slow wind (active region, Quiet Sun, helmet streamers) sources suggesting the slow wind, as defined by ion ratios, is formed across various sources.
The variability in the compositional makeup of the slower speed wind suggests many distinct solar sources.

The helium abundance (\ahe) is the hydrogen-to-helium number density ratio.
Often, it is expressed in units of percent \citep{Aellig:Ahe,Kasper:Ahe,Kasper:Ahe:Qstate,Alterman2019,Alterman2021,McIntosh:Ahe}
\ahe\ exhibits the most extreme values in coronal mass ejections, sometimes reaching over 20\% \citep{Song2020, Johnson2024}. Helium is generally in the form of He$^{2+}$ after leaving the corona, however occasionally measurements in the solar wind measure significant amount of He$^{+}$ that is believed to have originated at the Sun \citep{Rivera2020}.
Categorizing \ahe\ by \vsw\ provides more nuanced insight.

Broadly, solar wind helium observations from the Wind spacecraft aggregated across several solar cycles show helium abundances gradually increase with increasing solar wind speed up to \kms[\sim 400] and then saturates to $\sim 4\%$ \citep{Aellig:Ahe,Wind:SWE:bimax,Alterman2019,Yogesh2021}.
We define the solar wind speed as the hydrogen bulk speed.
This dependence of \ahe\ on \vsw\ is stronger in slower solar wind and during solar activity minima \citep{Aellig:Ahe}.
It also varies with heliographic latitude and, accounting for this heliographic variability, the gradient of \ahe\ as a function of \vsw\ in slow wind is linear \citep{Kasper:Ahe}.
Both the heliographic and \vsw\ dependencies are absent during solar maxima, suggesting that slow solar wind helium has two distinct sources: the streamer belt and active regions (ARs) \citep{Kasper:Ahe}.
Furthermore, the speed at which helium vanishes from the solar wind is \vv[259 \pm 12], which is within $1\sigma$ of the minimum observed solar wind speed, and may be related to how helium interacts with solar wind acceleration \citep{Kasper:Ahe} and is robust to analysis across multiple solar activity cycles \citep{Alterman2019}.
Extending the analysis of \ahe\ to multiple solar cycles shows that \ahe's variability with solar activity carries a \vsw-dependent phase lag \citep{Feldman1978,Alterman2019}, which is likely driven by changes in distinct slow solar wind source regions, i.e.~helmet streamers and ARs.
\ahe's \vsw-dependence also shows a rapid depletion or ``shutoff'' approximately \days[250] prior to solar minima across multiple solar activity cycles, which may be related to changes in the magnetic topology of solar wind source regions \citep{Alterman2021}.

Heavier elements ($Z>2$) observed in the ecliptic at \au[1] also express speed and solar cycle dependence which can provide additional insight to the transitional boundaries observed with He \citep{ACE:SWICS:AUX}.
Combining \ahe\ with $\C[6]/\C[5]$ and $\Ox[7]/\Ox[6]$ ratios shows that the temperature of slow wind solar wind source regions also varies with solar activity, decreasing with decreasing sunspot number (\SSN) \citep{Kasper:Ahe:Qstate}.
Extending the analysis to \AbSW[\Fe][\Ox], \citet{McIntosh:Ahe} infer that a decrease in plasma heating deep in the solar atmosphere during solar minima drives the decrease in \ahe\ with decreasing solar activity.
Extending this abundance analysis to additional element ratios and using \Hy-normalized abundances (\AbSW), \citet{ACE:SWICS:AUX} show \AbSW\ vary with solar activity in the same manner as heavy ion charge state ratios for both fast and slow solar wind, which further ties the variability of solar wind observations at \au[1] to changes in solar wind source region properties.
Clearly, examining a range of elements of a large range of chemical properties (e.g.~mass and FIP) will enable a more rigorous characterization of source region and solar wind release because those properties impact a given element's interaction with source region and solar wind release processes.

In this work, we extend the analysis of \citet{Aellig:Ahe,Kasper:Ahe,Kasper:Ahe:Qstate,Pilleri2015,Alterman2019,Alterman2021} to examine the dependence of heavy ion composition on solar wind speed using \emph{Advanced Composition Explorer's} \citep[ACE;][]{ACE} \emph{Solar Wind Ion Composition Spectrometer} \citep[SWICS;][]{ACE:SWICS} over the years 1998 to 2011.
To enhance our confidence in these results, we compare our observations of the helium abundance observed by ACE/SWICS to the same abundance observed by the Faraday cups that are part of the \emph{Wind} spacecraft's \emph{Solar Wind Experiment} \citep[SWE;][]{Wind:SWE}.
This paper proceeds as follows.
\sect{obs} briefly summarizes our observations.
\sect{analysis} presents our analysis.
In \sect{disc}, we discuss the results.
\sect{conclusion} then concludes.

\section{Observations \label{sec:obs}} 

We use data from both a Faraday cup (FC) and a mass spectrometer. 
They are on distinct instruments, both at the \nst{1} Earth-Sun Lagrange point (L1), but onboard different spacecraft.
Section 2.1 of \citet{LR:MSSW} summarizes the classes of instruments in detail.
Here, we identify the specific instruments, datasets, and data selection used.

\subsection{Wind/SWE Faraday Cup Observations}
We use observations of the solar wind speed (\vsw), hydrogen density, and helium density provided by the Wind \citep{Wind} Solar Wind Experiment \citep[SWE,][]{Wind:SWE} Faraday cups (FCs).
These observations are provided by CDAWeb at the native $\sim92$s cadence.
There are multiple SWE data sets, each each utilizing non-linear fitting to extract physical parameters, and optimized for different objectives \citep{Wind:SWE:bimax,Wind:SWE:dvapbimax,Alterman2018}
We use the data optimized for deriving the helium abundance, which is described by \citet{Wind:SWE:bimax,KasperThesis}.
This dataset only considers one proton population, which is nominally the proton core \citep{Alterman2018}.
Our data selection follows \citet{Alterman2019,Alterman2021}.

\subsection{ACE/SWICS Observations}
The Advanced Composition Explore \citep[ACE,][]{ACE} Solar Wind Ion Composition Spectrometer \citep[SWICS,][]{ACE:SWICS} is an energy -- time-of-flight mass spectrometer that provides heavy ion composition observations of H, He, C, N, O, Ne, Mg, Si, S, Fe at 2hr cadence \citep{ACE:SWICS:MLE}.
For ACE observations of \Hy\ and \He, we utilize data from the auxiliary channel of ACE/SWICS \citep{ACE:SWICS:AUX}.
For ACE observations of \Hy, we utilize data from the auxiliary channel of ACE/SWICS, a separate ESA channel of the instrument that is optimized for \Hy\ observations \citep{ACE:SWICS:AUX}.
Density and velocity are calculated from the first and second moments and then quality filtered, effectively eliminating contamination from the small amount of \He\ present.
We limit our study to the years 1998 to 2011, i.e. data from before SWICS' detector anomaly \citep{Zurbuchen2016}.
To isolate ambient and non-transient solar wind, we remove interplanetary coronal mass ejections \citep[ICMEs;][]{Richardson2010} and corotating interaction regions \citep[CIRs;][]{Mason2012b}.

\section{Analysis \label{sec:analysis}}

\subsection{The Fast/Slow Transition \label{sec:fs-transition}} 

\plotVswAb
The categorization of solar wind by speed into ``fast'' and ``slow'' can be considered overly broad and requires both reconsideration and additional detail. 
In this section, we utilize observations over the full range of \vsw\ observed by Wind/SWE to characterize the speed at which the helium abundances changes from characteristically ``slow'' to characteristically ``fast''.

\cref{fig:VswAb} the SWE helium abundance as a function of the proton speed observed at \emph{Wind}. 
The abundance is normalized to its photospheric value \citep{Asplund2021}.
The proton speed is binned in 45 quantiles over the range 250 to \kms[800].
Visual inspection shows that there is a point around \kms[\sim400] where the slope of \AbSW[\He] as a function of \vsw\ decreases.
Because quantiles divide the data into intervals with an equal number of observations and slower speeds are observed more frequently than faster speeds, quantiles provide greater resolution around the saturation speed \vs\ than fixed width intervals would.
We have normalized the observations in each column to the column's maximum value so that the overall trend of \AbSW[\He] with \vsw\ is not obscured by the \vsw\ sample frequency and this point is easily seen on inspection.

To quantify this transition, we have fit the trend of these distributions with the bi-linear function
\eqBilinear
$A(v)$ is the abundance normalized to its photospheric value
\begin{equation}
A = \left(X/\Hy\right) \! : \! \left(X/\Hy\right)_\mathrm{photo},
\end{equation}
where $X/\Hy = \n[X]/\n[\Hy]$ is the number density of species $X$ normalized to the \Hy\ number density.
For the two lines indicated by subscript 1 or 2, $A_i$ are the two different lines; $m_i$ are the slopes of the lines, and $v_i$ are the x-intercepts of the lines.
\citet{Kasper:Ahe} calls $v_i$ for the line in \cref{eq:bilinear} with the steeper slope (nominally the slow wind) the vanishing speed, i.e. the speed at which \ahe\ vanishes from the solar wind.
The two lines intersect at the saturation speed (\vs) where the abundances of both lines are equal $A_1 = A_2 = \As$.
The speed at the intersection between the two lines in \cref{eq:bilinear} is given by
\eqSspeed
As the intersection of two lines reduces the number of free parameters in the two equations to four, we choose to parameterize the fit function so that the free parameters are \vs, \As, $v_1$ (the x-intercept of the $v < \vs$ line), and $m_2$ (the slope of the $v > \vs$ line).
This parameterization directly quantifies the point at which \AbSW[\He] transitions from its slow to fast wind values (\vs, \As) and provides uncertainties for it.
To account for the variable frequency with which Wind samples different solar speeds and reduce the impact of extreme values of \AbSW[\He], we select data in bins within 80\% of the column maximum and then calculate the mean and standard deviation in log-space.
We then fit these mean values with the minimum of two lines \cref{eq:bilinear}.
Each column's standard deviation is used as the weight.
The green line and error bars in \cref{fig:VswAb} show the mean and standard deviation in each column.
The pink dash-dotted line shows the trend.
As \AbSW[\He] saturates to its fast wind value \As*[0.520 \pm 0.004] at speeds $v > \vs[399 \pm 2]$, we refer to this as the saturation speed \vs\ and the corresponding abundance \As\ as the saturation abundance.
We refer to this coordinate \satpoint\ as the ``saturation point''.
Although it has been suggested to refer to this point as a transition, that would imply the symbol $v_t$, which could easily be confused with the thermal speed.

\plotVswAbComparison*
Solar wind acceleration and heating leave signatures in the trends of heavy ions observed at \au[1].
To characterize the impact of these signatures on the transition between ``slow'' and ``fast'' wind, we have examined the SWICS abundances of \He, \C, \N, \Ox, \Ne, \N, \Mg, \Si, \Su, and \Fe\ all normalized to hydrogen $(\AbSW[X])$ in the same manner as \AbSW[\He] observed by SWE in \cref{fig:VswAb}.
\cref{fig:VswAbComparison} plots these SWICS and SWE abundances as a function of \vsw\ observed at their respective spacecraft.
We limit the observations plotted to those below \kms[600] because \cref{fig:VswAb} shows the transition is at speeds more than \kms[100] slower and due to large scatter at higher speeds.
Every second data point is marked.
Labels on the right side of the plot indicate the species, with \emph{SWE} indicating \AbSW[\He] observed by Wind/SWE.

\plotVswAbComparisonScaled*
We have fit the plotted observations with the bi-linear function in \cref{eq:bilinear}.
Vertical, semi-transparent lines in \cref{fig:VswAbComparison} indicate the speeds at which the slope of the bi-linear fit changes, i.e. each species' saturation speed.
\cref{tbl:saturation} gives saturation speeds (\vs) and abundances (\As) along with the other parameters for the fits and the percentage of the observations in the $v < \vs$ regime.
This latter quantity shows that the slow wind portion of each species contains a non-trivial fraction of that species' observations.
Although there is more scatter in the plots due to SWICS' lower time resolution and the time period over which observations are available is smaller, they all show a fast-to-slow transition.
Excluding \He, the average heavy ion saturation speed is \vs[327 \pm 2].
We have performed a similar average for the low and high FIP elements and those \vs\ are the same to with in the propagated uncertainties.
Broadly, all species show similar qualitative behavior in that all \AbSW\ monotonically increase with increasing \vsw\ and have a different gradients at speeds above and below their respective \vs.

\TblSaturation
To better characterize these gradients, \cref{fig:VswAbComparison:scaled} re-plots the data in \cref{fig:VswAbComparison} and scales the observations to the saturation point $(\vs, \As)$, which is plotted at $(1, 1)$.
\He\ for SWE and SWICS do not extend to as large of a value on the x-axis as other species because $v_{s,\He}$ is larger than $v_{s,\mathrm{Heavy}}$.
This figure shows that below the saturation speed ($v < \vs$), the gradient of scaled abundances as a function of \vsw\ is indistinguishable across the different species.
For speeds $v > \vs$, each species' gradient is shallower than its $v < \vs$ gradient and these gradients are different for the different species.
\C, \N, and \Ox\ have the steepest gradients that are most similar to their $v < \vs$ gradient.
\Fe\ and \He\ have the shallowest gradients that are most distinct from their $v < \vs$ gradient.
\Ne, \Mg, \Si, and \Su\ are the intermediate case.

\PlotVsChemistry
\subsection{Saturation Properties \label{sec:saturation}} 
In this section, we characterize the saturation of each species at its $(\vs, \As)$ point.
The figures \cref{fig:vs-chemistry,fig:as-chemistry,fig:af-chemistry} use a consistent style in which each species' color and marker match its style in \cref{fig:VswAbComparison,fig:VswAbComparison:scaled}.
Data points are connected with a black dotted line to aid the eye.
The top axis indicates the species.
In the case of \AbSW[\He] from SWE, the marker is explicitly labeled to differentiate it from SWICS' \AbSW[\He].

\cref{fig:vs-chemistry} plots \vs\ as a function of \FIP.
The vertical dashed line is \eV[11], the nominal change between high and low \FIP\ \citep{STQT:selection-abundances}.
It shows that \vs\ for all heavy ions are within the 1$\sigma$ fit uncertainty of each other.
\Ne\ is the exception in that the upper range of $v_{s,\Ne}$ is slower than the lower range of $v_{s,\Si}$.
Excluding \He, the average saturation speed for heavier elements is \vs[327 \pm 2].
The \AbSW[\He] saturation speed observed by SWICS is \vs[390 \pm 4].
For comparison, \vs\ for \AbSW[\He] observed by SWE is \vs[399 \pm 2], a 3 to \kms[15] difference, which we consider small in comparison to the \kms[63 \pm 4.5] difference between \vs\ for SWICS' \AbSW[\He] and heavier elements.

\PlotAsChemistry
\cref{fig:as-chemistry} plots the saturation abundance \As\ as a function of \FIP.
Again, the vertical dashed line indicates \eV[11], the nominal transition between low and high \FIP.
This figure shows the expected rend that low \FIP\ elements (\FIP[11][<]) are enhanced more than high \FIP\ elements by a factor of approximately 2.
This trend is expected from \citet{Zurbuchen2016}, who compared \AbSW[X][\Ox] in interplanetary coronal mass ejections (ICMEs), fast solar wind, and slow solar wind.
That \Su\ behaves like a low \FIP\ element is also consistent with observations of suprathermal ions during quiet times \citep{STQT:selection-abundances}.

\PlotFastestMass
To characterize the change in gradients for speeds $v > \vs$, we have normalized the abundance at \kms[592], which is the fastest considered in this analysis and are filled markers in \cref{fig:VswAbComparison,fig:VswAbComparison:scaled}, to \As\ and plotted it as a function of element mass (\M) in \cref{fig:af-chemistry}.
Normalizing to \As\ removes the photospheric normalization.
We choose \M\ because these quantities are not organized by \FIP\ and several \M-dependent fractionation processes have been observed \citep{Rivera2021,Lepri2021,Pilleri2015,Weberg2012,Wurz2000}.
\cref{tbl:saturation} includes the abundances at \kms[592] under $A\left(\kms[592]\right)$.
Normalizing this fastest abundance to \As\ allows us to qualitatively characterize the gradient in a manner that accounts for the known differences in heavy ion abundances due to fractionation processes in the chromosphere and transition region \citep{Laming2004,Laming2009,LR:FIP,Schwadron1999,Geiss1982a,Geiss1995b}.
Error bars are the propagated errors including the fit uncertainty in \As\ and the statistical uncertainty in the \kms[592] data point.
Excluding \He, which is a known outlier in comparison to heavier elements, we observe a roughly monotonic decreases in $A(\kms[592])/\As$ with three clusters.
\C, \N, and \Ox\ are clustered in the range $\sim 1.8$ to 2.
\Mg, \Ne, \Si, and \Su\ are clustered at $\sim 1.5$.
\Fe's $A(\kms[592])/\As$ is approximately 1.2, which is comparable to \He's.

\section{Discussion \label{sec:disc}}

Broadly, the solar wind can be classified into two types based on its speed: fast and slow.
The difference between fast and slow wind is often chosen \emph{ad hoc} to be somewhere in the range of 400 to \kms[600].
For example, an often cited justification for classifying the solar wind as fast or slow is that the helium abundance has a strong, positive gradient with solar wind speed in slow wind and saturates to a fixed value in fast wind. 
We identify the speed-abundance pair at which a given element transitions from slow wind-like to fast wind-like behavior as $\left(\vs,\As\right)$ by fitting \cref{eq:bilinear} to the trends of \AbSW\ as a function of \vsw.
Using several solar cycles of observations from the Wind Faraday cups, \cref{fig:VswAb} shows that, statistically, \AbSW[\He] saturates to a photospheric-normalized abundance of \As*[0.520 \pm 0.004] at \vs[399 \pm 2].
However, \emph{in situ} observations of kinetic properties \citep{Kasper2008,Kasper2017,Tracy2016,Alterman2018,Klein2018,Martinovic2019,Martinovic2021a}, chemical makeup and charge state properties \citep{vonSteiger2000,Geiss1995,Geiss1995b,Zhao:InSituComposition:Sources,Zhao2022,Xu2014,Fu2017,Fu2015}, and cross helicities \citep{Tu1995,LR:turbulence,DAmicis2021} indicate that the helium abundance alone carries insufficient information to fully characterize the transition between fast and slow solar wind.

We repeat the analysis in \cref{fig:VswAb} for all heavy ion abundances \AbSW\ observed by ACE/SWICS, both \He\ and heavier elements.
\cref{fig:VswAbComparison} plots the observations.
\cref{fig:vs-chemistry} summarizes the derived saturation speeds \vs\ as a function of \FIP.
In the case of SWICS' \AbSW[\He], the abundance saturates to it's fast wind value at \vs[390 \pm 3], which is at most \kms[14] slower than \vs\ observed by Wind/SWE.
\cref{fig:vs-chemistry} shows that, with the exception of \Ne\ and \Si, heavy element \vs\ are all within their mutual uncertainties.
As such, we take their weighted mean \vs[327 \pm 2] as the typical heavy element saturation speed, which is indicated by a horizontal blue line in \cref{fig:vs-chemistry}.
This heavy ion \vs\ is \kms[63 \pm 5] slower than \vs\ observed by SWICS, which is $7\times$ larger than the difference between \vs\ observed by SWICS and SWE for \AbSW[\He].
Given that SWICS is a time-of-flight mass spectrometer, SWE consist of two Faraday cups, and these two instruments are mounted on different spacecraft, a difference of $< 4\%$ between \vs\ derived form SWICS and SWE measurements seems negligible in comparison to the difference between \vs\ for \He\ and heavier elements observed by SWICS.
This suggests that SWICS and SWE observations of \AbSW[\He] are statistically consistent over the long duration of the observations used in this study and provides high confidence that there is not a systematic difference between these instruments that is significant on the scales statistically analyzed in this work.
As such, we use the SWICS \He\ observations for comparison with heavier element abundances.

\cref{fig:as-chemistry} plots the saturation abundance \As, the abundances at \vs, as a function of \FIP\ and shows the expected dependence.
In qualitative with \citet{Zurbuchen2016,VonSteiger2016}, low \FIP\ elements are enhanced from their photospheric values by $\sim2\times$ more than high \FIP\ elements.
The agreement is qualitative because \citet{Zurbuchen2016,VonSteiger2016} analyze \AbSW[X][\Ox] or normalize their \AbSW\ values to fast wind \AbSW, not photospheric values.
The observed \FIP-dependence suggests the unsurprising result that the abundances characteristic of the transition between slow and fast solar wind are driven in chromosphere, where solar wind abundances are fractionated by the pondermotive force \citep{Laming2004,Laming2009,LR:FIP,Schwadron1999,Geiss1982a,Geiss1995b}.
For completeness, we have also examined these saturation abundances as a function of element mass (\M) and typical solar wind charge state (\Q) \citep{vonSteiger1997,Desai2006b}.
Although beyond the scope of this paper and not shown for space, we note that \As\ for high \FIP\ elements shows a monotonic decrease with increasing \M\ and \Q.

To contextualize the gradients of \AbSW\ as a function of \vsw\ at speeds slower and faster than \vs, \cref{fig:VswAbComparison:scaled} scales the observations plotted in \cref{fig:VswAbComparison} to each species' transition point $\left(\vs,\As\right)$.
As abundance are set by \FIP\ in the chromosphere and the solar wind's asymptotic speed is set above this height, such a normalization removes any (simple) offsets that are due to preferential element or ion coupling to these mechanisms and reveals any trends obscured by them.
This shows that in solar wind with speeds $v < \vs$, the gradients of \AbSW\ as a function of \vsw\ are effectively indistinguishable.
This is not the case for speeds $v > \vs$, for which there may be three distinct groups.
For high \FIP\ elements heavier than \He\ (i.e.~\C, \N, and \Ox), the change in gradient at $v > \vs$ is least significant.
Low \FIP\ elements \Mg, \Si, and \Su\ have an intermediate change in gradient.
\Ne\ and \Fe\ are the exceptions to this trend.
Although \Ne\ is high \FIP, the change in its gradient is more similar to the low \FIP\ elements than other high \FIP\ elements.
In the case of \Fe, its gradient at $v > \vs$ is most similar to \He, which is generally an exception to composition trends.
\cref{fig:af-chemistry} plots the fastest reported abundance at \kms[592], the fastest speed plotted in \cref{fig:VswAbComparison} and indicated by filled markers, normalized to \As\ as a function of mass and shows that, with the exception of \He, these transition abundances \As\ are well-ordered by mass, indicating a possible mass-dependent fractionation process in solar wind with $v > \vs$.

\subsection{Implications for the Fast/Slow Solar Wind Transition and Solar Wind Sources \label{sec:disc:fs-transition}}

Coronal conditions are such that energy conversion at the sonic point is insufficient to yield the asymptotic fast wind speeds observed at \au[1] \citep{Leer1980,Hansteen2012}.
Rather, additional energy must be supplied to the solar wind to achieve the asymptotically fastest observed non-transient solar wind.
In contrast to a quantity that evolves with distance like speed, elemental abundances are conserved quantities.
This is key to utilizing \FIP\ fractionation as an \emph{in situ} diagnostic of solar wind source regions at the Sun.
The combined measurements of abundances and speed therefore probe a combination of solar wind source region and transport effects.

The overall trends of \AbSW\ with \vsw\ that show two distinct gradients at speeds $< \vs$ and $> \vs$ in \cref{fig:VswAbComparison,fig:VswAbComparison:scaled} are consistent with the two-state solar wind paradigm under which fast wind is from coronal holes (CH) with magnetic fields that are continuously open to heliosphere and slow wind is from equatorial sources with more complex, likely intermittently open magnetic topologies.
The difference between low and high \FIP\ \As\ in \cref{fig:as-chemistry} is consistent with Figure 5 in \citet{Zurbuchen2016} and a \FIP-dependent process at the Sun fractionating elemental abundances in the chromosphere, which suggests that the fast/slow transition is independent of the \FIP\ effect.
Given that \vs\ is similar for all elements heavier than \He\ and the solar wind speed is set above heights where \FIP\ fractionation occurs, we make the unsurprising inference that the transition between fast and slow solar wind sources occurs at heights above where the pondermotive force or any similar process that induces fractionation impacts the solar plasma.
The difference in \He\ \vs\ and heavy element \vs\ along with the difference in gradients of \AbSW\ at speeds below and above \vs\ require a more nuanced interpretation.

Two critical distances associated with solar wind acceleration are the sonic and Alfvén critical point, which we denote by \rc\ and \rA, respectively.
The sonic point is the distance from the Sun's surface at which the solar wind's bulk speed exceeds the thermal speed. Under the Parker model \citep{Parker1958a}, this happens when the plasma's thermal energy is converted to kinetic energy and the solar wind becomes supersonic.
The Alfvén point or surface is the distance at which the solar wind's speed exceeds the local Alfvén speed, i.e.~the solar wind is traveling faster than information can propagate along magnetic field lines attached to the Sun's surface.
\citet{PSP:SWEAP:Ra} observed the Alfvén surface to be just below \Rs[20].
Alfvén waves are a possible source of the energy above either \rc\ and/or \rA\ necessary for the solar wind to achieve its asymptotic fast wind values.

Our analysis assumes that there exists a characteristic point $\left(\vs,\As\right)$ for each species' abundance as a function of \vsw\ and this point statistically indicates a transition between measurements of plasma from CH and equatorial sources that each have characteristic speeds, abundances, and abundance gradients as a function of \vsw.
By normalizing the observed trends in \cref{fig:VswAbComparison} to this point $\left(\vs,\As\right)$, \cref{fig:VswAbComparison:scaled} accounts for any offsets present in these trends that, by assumption, are unrelated to the process(es) that lead to the two different gradients above and below \vs.
Because the gradients for $v < \vs$ are consistent across species, we infer that there is no process that preferentially couples to and drives changes in any one species abundances or as a function of a element properties like \M, \Q, \MpQ, or \FIP.
In other words, there is no fractionation process in the slow wind beyond that which is introduced by the \FIP\ effect in the chromosphere, as observed by the vertical scaling in \cref{fig:VswAbComparison}.

\plotHingeVswComparison
To contextualize the difference in \vs\ for \He\ and heavier elements, \cref{fig:xh-vsw-comparison} plots these speeds with the probability density of \vsw\ observations from SWICS.
Vertical green lines indicate \vs\ for \He\ observed by SWICS and the weighted average of \vs\ for heavier elements; line widths are the range of values covered by the uncertainties.
The vertical dotted line is the peak of the \vsw\ distribution.
This visualization clearly shows that $\vs_\He_ > \vs_\mathrm{Heavy}_$ and these two characteristic speeds are separated by the solar wind distribution's peak.
Using SWE observations of \vsw\ and \AbSW[\He] does not change this interpretation.
That \vs\ for elements heavier than \He\ are mutually consistent further suggests that there is no process that preferentially accelerates heavier elements at distances from the Sun $> \rc$ or $> \rA$, where processes that occur during transport continue to accelerate the fast solar wind to its asymptotic values.
However, this does not rule out such an \emph{in situ} acceleration process impacting \He\ for $v > \vs$. 
For example, \He's large density with respect to heavier elements makes it more likely to be impacted by beam instabilities, inter-particle Coulomb collisions, and Alfvén wave transport.
If the difference between \vs_\He_\ and \vs_\heavy_ is due to \emph{in situ} an acceleration process at distances from the Sun above the sonic point \rc\ preferentially coupling to \He\ in comparison to heavier elements, the difference between \vs_\He_ and \vs_\heavy_ has profound implications for the definitions of fast and slow solar wind.

The helium abundance is a key motivator for associating fast and slow solar wind to with distinct solar sources.
The scaled observations in \cref{fig:VswAbComparison:scaled} show that $\AbSW\left(v < \vs\right)$ have indistinguishable gradients.
\cref{fig:VswAbComparison,fig:as-chemistry} show the expected \FIP\ fractionation.
From this, we infer that observations from $v < \vs$ are equatorial in origin and observations from $v > \vs$ are CH in origin.
A consequence of this that differentiating between fast and slow solar wind with a threshold in the range of 400 to \kms[600] would yield ``slow wind'' abundances with chemical compositions reflecting a mixture of source regions continuously magnetically open to the heliosphere (e.g.~CHs) and those that are only intermittently open (i.e.~equatorial sources).
Such a mixing would obscure our ability to properly map slow solar wind back to its solar origin.

\subsection{Implications for Fast Wind Fractionation and Solar Wind Acceleration \label{sec:disc:fractionation-acceleration}} 

\citet{Pilleri2015} study heavy ion abundances normalized to \Mg\ during solar minimum 23 and maximum 24 using ACE/SWICS data to contextualize \emph{Genesis} observations \citep{Genesis}.
They divide their observations into those from CHs, equatorial sources, and CMEs; calculate their abundances with respect to magnesium; and include an analysis of \AbSW[X][\Mg]'s dependence on \vsw.
They divide their solar wind observations into CH and equatorial in origin based on a change in solar wind speed.
They report mass-dependent fractionation trends comparable to ours where by \AbSW[\Ox][\Mg], \AbSW[\C][\Mg], and \AbSW[\He][\Mg] show two distinct gradients above and below \kms[\sim400], which roughly divides equatorial and CH wind.
That their threshold speed is faster than \vs\ we report for heavy ions is unsurprising because they separate CH and equatorial solar wind by following \citet{Reisenfeld2013} and setting a threshold on solar wind speed at stream interfaces that is \kms[425] for rarefaction regions and \kms[525] compression regions.
These authors also attribute their fractionation to a secondary dependence of the pondermotive force predicted by \citet{Laming2004}.
However, \citet{Laming2004,Laming2009,LR:FIP} emphasize that the pondermotive force is effectively mass-independent.
As such, another explanation may be necessary.

One possibility is that the solar wind speed varies between the center and boundary of CHs \citep{Zhao:InSituComposition:Sources}.
Performing a superposed epoch analysis of 66 Carrington rotation long intervals of CH solar wind, \citet{Borovsky2016} shows a gradient of speed in time over the range of 400 to \kms[600], which is the range over which we observe an enhancement in $A\left(\kms[592]\right)/\As$.
However, \citet{Borovsky2016} also shows that \AbSW[\Fe][\Ox] does not vary over these intervals.
As the abundance normalized to \As\ of \AbSW[\Fe] has the shallowest gradient for speeds $v > \vs$ and that of \AbSW[\Ox] has one of the strongest gradients, this suggests that our observed mass fractionation is not a result of position within a CH or distance from its edge.

In the case of Coulomb friction with \Hy\ dragging heavy ions out of the corona, \citet{Bodmer1998a,Bodmer2000} show that there is a mass dependence and the associated fractionation would be stronger in slow than fast wind.
Such a trend does not agree with the fractionation we observe in fast wind and lack of fractionation in slow wind when \FIP\ fractionation is accounted for by normalizing abundances to \As.
As such, we rule out Coulomb friction as a source of the observed $v > \vs$ trend.

\citet{Rivera2021} report signatures of mass-dependent fractionation in CMEs in which heavy ion abundances decrease with increasing mass when absolute CME abundances (\AbSW) are normalized to ambient absolute solar wind abundances.
\citet{Lepri2021} report a similar trend for prominence material, though of higher values.
\citet{Rivera2021} attribute this trend to gravitational settling.
\citet{Weberg2012} also demonstrate that gravitational settling leads to mass-dependent fractionation in ambient solar wind.
However, gravitational settling has a timescale on the order of days and requires closed loops, which are common in equatorial regions where slow wind originates, not CH regions from which fast wind is from.
As such, the mass-dependent fractionation observed in solar wind with speeds $v > \vs$ is also unlikely due to gravitational settling.

In short, we have shown that the observed fast wind enhancements of heavy ion abundances above \As\ and the corresponding mass-dependent fractionation are inconsistent with the effects of gravitational settling, \Hy\ dragging coronal heavy ions into the solar wind by means of Coulomb friction, and gradients across CHs.
\cite{Pilleri2015} suggest that mass-dependent fractionation is consistent with a pondermotive-driven \FIP\ effect.
Although our trends qualitatively agree with \citet{Pilleri2015}, \citet{Laming2004,Laming2009,LR:FIP} explicitly state that a pondermotive-driven \FIP\ effect is mass independent.
As such, this may be novel mass-dependent, fast wind fractionation or the fractionation depends on a different quantity.

\PlotFastestQState
In addition to \FIP\ and \M, the elements reported here are summed over a series of charge states and, as such, have an average charge state.
\citet[Eq.~(14)]{Renaud2010} propose a robust coefficient of determination (\rsqw) that is appropriate for rapidly determining if a model reasonably fits a set of data that includes uncertainties.
Excluding \He, we have fit the fastest abundances observed $A\left(\kms[592]\right)/\As$ as a function of \FIP, \M, solar wind charge state \Q\ \citep{vonSteiger1997,Desai2006b}, \MpQ, and $M^2/Q^2$ with a line and calculated \rsqw\ as a simple means of quantifying how well-organized $A\left(\kms[592]\right)/\As$ is by each quantity.
All \rsqw\ are $< 0.55$ except for the dependence on average charge state, for which \rsqw[0.95].
\cref{fig:af-qstate} plots $A\left(\kms[592]\right)/\As$ as a function of charge state in the style of \cref{fig:vs-chemistry,fig:af-chemistry}.
Although we do not have an explanation for this result, it suggests that the fractionation may depend on solar wind charge state.

Beyond the reported \M- or \Q-dependent fractionation of $A\left(v > \vs\right)/\As$, \cref{fig:VswAbComparison:scaled} shows that the degree of fractionation increases with \vsw\ for $v > \vs$.
To achieve the asymptotically fastest speeds observed at \au[1], energy must be deposited into the solar wind above the sonic critical point \rc\ \citep{Leer1980,Hansteen2012}.
The faster the speed before this yet-to-be-identified mechanism accelerates the solar wind and the more energy deposited by it, the larger the asymptotic speed.
Recent work \citep{Rivera2024,Bale2023,Raouafi2023} shows that the solar wind's acceleration at distances $r > \rA$ is driven by the deposition of energy into the solar wind from switchbacks dissipation during solar wind propagation through interplanetary space.
Given all ions are observed at fast wind speeds that the solar wind requires such energy deposition to reach, the increase in the degree of heavy ion fractionation may indicate that there is a preferential coupling between these heavy ions and the energy deposition process.
On the other hand, we have argued that the consistency of \vs\ across the heavy ions suggests that $\vs_\He_ > \vs_\heavy_$ may indicate \He\ is impacted by this yet-to-be-identified acceleration mechanism at distances from the sun above \rc\ and heavier elements are not.
If this is the case, then the dependence of the fractionation process for $v > \vs$ must be located at or near \rc.

\section{Conclusion \label{sec:conclusion}} 
Under the two-state paradigm, the solar wind is classified into fast and slow based on whether its speed is above or below a threshold value.
This threshold is typically between $\sim400$ and \kms[\sim600] and chosen in an \emph{ad hoc} or heuristic fashion.
Fast solar wind with speeds above this threshold value are typically observed to come from  magnetically open, typically polar regions like coronal holes (CHs) \citep{Phillips1994,Geiss1995}.
Slow solar wind is from more equatorial regions with magnetic fields that may only be intermittently open to the heliosphere \citep{Fisk1999,Subramanian2010,Antiochos2011,Crooker2012,Abbo2016,Antonucci2005}.
Analysis of the solar wind's kinetic \citep{Kasper2008,Kasper2017,Tracy2016,Wind:SWE:bimax,Fu2018,Stakhiv2016,Alterman2018},  heavy ion abundance along with charge-state ratio \citep{vonSteiger2000,Geiss1995,Geiss1995b,Zhao:InSituComposition:Sources,Zhao2022,Xu2014,Fu2017,Fu2015}, Alfvénicity \citep{DAmicis2021a,DAmicis2021,LR:turbulence,Tu1995}, and heavy ion composition in switchbacks \citep{Rivera2024a} provide a more nuanced picture in which there are multiple classes of solar wind.

Motivated by the distinct gradients of \AbSW[\He] as a function of \vsw\ observed by Wind/SWE above and below the speed \vs, we have investigated the variation of \AbSW\ observed by ACE/SWICS as a function of \vsw.
We have made the following observations and inferences.
\begin{enumerate}
\item All species have two distinct gradients as a function of \vsw\ and these gradients are shallower above the speed \vs.
From this, we infer that the change in gradients is a signature of differences in the magnetic topology at distinct types of solar wind source regions.
\item The \He\ saturation speed is \vs[399 \pm 2] (observed by SWE) and \vs[390 \pm 4] (observed by SWICS).
We interpret this as showing that SWE and SWICS \AbSW[\He] are statistically consistent over the years 1998 to 2011.
\item The average \vs\ across elements heavier than \He\ is \vs[327 \pm 2], independent of species, which is \kms[63 \pm 4.5] slower than \vs\ for SWICS' \AbSW[\He].
Moreover, the speed \vs\ for heavy elements is slower than the peak of the solar wind distribution and the speed \vs\ for \He\ is faster than the peak of the solar wind distribution.
From this, we infer that \He\ may be impacted by the acceleration at heights above the sonic point that is necessary for non-transient solar wind to reach the asymptotically fastest speeds observed at \au[1] and heavy elements are not.
\item If our inferences about the change in gradients of \AbSW\ as a function of \vsw\ across \vs\ and the observation that $\vs_\He_ > \vs_\heavy_$ hold, then this implies that setting a threshold for differentiating between slow and fast solar wind in the range of 400 to \kms[600] may lead to a ``slow'' solar wind with a chemical makeup that is a mixture of solar wind from CH and equatorial regions that are only intermittently open to the heliosphere.
\item The saturation abundances \As\ at the speed \vs\ are ordered by \FIP\ and show an expected fractionation pattern.
From this, we unsurprisingly infer that the fast/slow solar wind transition is the result of a mechanism that impacts the solar wind at heights above chromosphere, above where processes like the pondermotive force that would fractionates the solar plasma occur.
\item When normalized to the point $\left(\vs,\As\right)$, the gradients for elements heavier than \He\ are indistinguishable for $v < \vs$.
We interpret this observation as a signature that there is not a mechanism preferentially coupled to and driving slow wind abundance gradients as a function of species.
\item When normalized to the point $\left(\vs,\As\right)$, the gradients for elements heavier than \He\ are indistinguishable for $v > \vs$ are ordered by \M\ or average solar wind \Q\ for $v > \vs$.
Although average charge state provides a better ordering of the ratio of the fastest reported abundances to the saturation abundances $A\left(\kms[592]\right)/\As$ as indicated by the weighted coefficient of determination, such a fractionation process that only depends on average charge state is difficult to justify.
Even though we have ruled out multiple mass-dependent mechanisms as possible sources of the observed fractionation at speeds $v > \vs$, this leaves such a charge state dependent fractionation unsatisfying.
\end{enumerate}

The bimodal nature of the solar wind's distribution is most pronounced during solar minima when coronal holes are restricted to the Sun's polar regions and its equatorial regions are dominated by helmet streamers, pseudostreamers, and other features with magnetic topologies that are not connected to the heliosphere in a simple, radial fashion.
Furthermore, additional properties like the solar wind's Alfvénicity have shown that there is solar wind with speeds that are traditionally considered to be slow, but fast wind kinetic, chemical, and charge state properties.
This Alfvénic slow wind believed to emanate from coronal holes and not equatorial sources.
Further analysis of the solar wind's chemical makeup and its variation as a function of Alfvénicity and solar activity should provide additional insight into the relationship between \emph{in situ} solar wind observations and their sources on the Sun.

\begin{acknowledgements}
The authors thank Ruth Skokie for discussions of ACE/SWEPAM data.
BLA acknowledges NASA Grants 80NSSC22K0645 (LWS/TM), 80NSSC22K1011 (LWS), and 80NSSC20K1844.
YJR acknowledges support from the Future Faculty Leaders postdoctoral fellowship at Harvard University.
JMR and STL acknowledge NASA contract 80NSSC23K0542 (ACE/SWICS).
\end{acknowledgements}

\bibliography{Zotero.bib}{}

\begin{thebibliography}{117}
\expandafter\ifx\csname natexlab\endcsname\relax\def\natexlab#1{#1}\fi

\bibitem[{Abbo {et~al.}(2016)Abbo, Ofman, Antiochos, Hansteen, Harra, Ko,
  Lapenta, Li, Riley, Strachan, von Steiger, \& Wang}]{Abbo2016}
Abbo, L., Ofman, L., Antiochos, S.~K., {et~al.} 2016, Space Science Reviews,
  201, 55, tex.ids= Abbo2016a

\bibitem[{Acuña {et~al.}(1995)Acuña, Ogilvie, Baker, Curtis, Fairfield, \&
  Mish}]{Wind}
Acuña, M.~H., Ogilvie, K.~W., Baker, D.~N., {et~al.} 1995, Space Science
  Reviews, 71, 5

\bibitem[{Aellig {et~al.}(2001)Aellig, Lazarus, \& Steinberg}]{Aellig:Ahe}
Aellig, M.~R., Lazarus, A.~J., \& Steinberg, J.~T. 2001, Geophysical Research
  Letters, 28, 2767

\bibitem[{Alterman {et~al.}(2023)Alterman, Desai, Dayeh, Mason, \&
  Ho}]{STQT:selection-abundances}
Alterman, B.~L., Desai, M.~I., Dayeh, M.~A., Mason, G.~M., \& Ho, G. 2023, The
  Astrophysical Journal, 952, 42

\bibitem[{Alterman \& Kasper(2019)}]{Alterman2019}
Alterman, B.~L. \& Kasper, J.~C. 2019, The Astrophysical Journal, 879, L6,
  publisher: IOP Publishing

\bibitem[{Alterman {et~al.}(2021)Alterman, Kasper, Leamon, \&
  McIntosh}]{Alterman2021}
Alterman, B.~L., Kasper, J.~C., Leamon, R.~J., \& McIntosh, S.~W. 2021, Solar
  Physics, 296, 67, arXiv: 2006.04669 Publisher: The Author(s), under exclusive
  licence to Springer Nature B.V. ISBN: 1120702101801

\bibitem[{Alterman {et~al.}(2018)Alterman, Kasper, Stevens, \&
  Koval}]{Alterman2018}
Alterman, B.~L., Kasper, J.~C., Stevens, M., \& Koval, A. 2018, The
  Astrophysical Journal, 864, 112, publisher: IOP Publishing

\bibitem[{Antiochos {et~al.}(2011)Antiochos, Mikic, Titov, Lionello, \&
  Linker}]{Antiochos2011}
Antiochos, S.~K., Mikic, Z., Titov, V.~S., Lionello, R., \& Linker, J.~A. 2011,
  The Astrophysical Journal, 112, arXiv: 1102.3704 tex.ids= Antiochos2011a

\bibitem[{Antonucci {et~al.}(2005)Antonucci, Abbo, \& Dodero}]{Antonucci2005}
Antonucci, E., Abbo, L., \& Dodero, M.~A. 2005, Astronomy \& Astrophysics, 435,
  699

\bibitem[{Asplund {et~al.}(2021)Asplund, Amarsi, \& Grevesse}]{Asplund2021}
Asplund, M., Amarsi, A.~M., \& Grevesse, N. 2021, Astronomy \& Astrophysics,
  653, A141

\bibitem[{Baker {et~al.}(2023)Baker, Démoulin, Yardley, Mihailescu, Van
  Driel-Gesztelyi, D’Amicis, Long, To, Owen, Horbury, Brooks, Perrone,
  French, James, Janvier, Matthews, Stangalini, Valori, Smith, Cuadrado, Peter,
  Schuehle, Harra, Barczynski, Berghmans, Zhukov, Rodriguez, \&
  Verbeeck}]{Baker2023}
Baker, D., Démoulin, P., Yardley, S.~L., {et~al.} 2023, The Astrophysical
  Journal, 950, 65

\bibitem[{Bale {et~al.}(2023)Bale, Drake, McManus, Desai, Badman, Larson,
  Swisdak, Horbury, Raouafi, Phan, Velli, McComas, Cohen, Mitchell, Panasenco,
  \& Kasper}]{Bale2023}
Bale, S.~D., Drake, J.~F., McManus, M.~D., {et~al.} 2023, Nature, 618, 252

\bibitem[{Bodmer \& Bochsler(1998)}]{Bodmer1998a}
Bodmer, R. \& Bochsler, P. 1998, Physics and Chemistry of the Earth, 23, 683

\bibitem[{Bodmer \& Bochsler(2000)}]{Bodmer2000}
Bodmer, R. \& Bochsler, P. 2000, Journal of Geophysical Research: Space
  Physics, 105, 47

\bibitem[{Borovsky(2016)}]{Borovsky2016}
Borovsky, J. 2016, Journal of Geophysical Research A: Space Physics, 121, 5055,
  iSBN: 2169-9402 tex.ids= Borovsky2016a, Borovsky2016b

\bibitem[{Brooks {et~al.}(2015)Brooks, Ugarte-Urra, \& Warren}]{Brooks2015}
Brooks, D.~H., Ugarte-Urra, I., \& Warren, H.~P. 2015, Nature Communications,
  6, publisher: Nature Publishing Group

\bibitem[{Bruno \& Carbone(2013)}]{LR:turbulence}
Bruno, R. \& Carbone, V. 2013, Living Reviews in Solar Physics, 10, 1

\bibitem[{Burnett {et~al.}(2003)Burnett, Barraclough, Bennett, Neugebauer,
  Oldham, Sasaki, Sevilla, Smith, Stansbery, Sweetnam, \& Wiens}]{Genesis}
Burnett, D.~S., Barraclough, B.~L., Bennett, R., {et~al.} 2003, Space Science
  Reviews, 105, 509, iSBN: 0038-6308

\bibitem[{Crooker {et~al.}(2012)Crooker, Antiochos, Zhao, \&
  Neugebauer}]{Crooker2012}
Crooker, N.~U., Antiochos, S.~K., Zhao, X., \& Neugebauer, M. 2012, Journal of
  Geophysical Research: Space Physics, 117, n/a, iSBN: 0148-0227

\bibitem[{D'Amicis \& Bruno(2015)}]{DAmicis2015}
D'Amicis, R. \& Bruno, R. 2015, Astrophysical Journal, 805, 1, publisher: IOP
  Publishing ISBN: 1538-4357

\bibitem[{Del~Zanna(2019)}]{DelZanna2019}
Del~Zanna, G. 2019, Astronomy \& Astrophysics, 624, A36

\bibitem[{Desai {et~al.}(2006)Desai, Mason, Gold, Krimigis, Cohen, Mewaldt,
  Mazur, \& Dwyer}]{Desai2006b}
Desai, M., Mason, G., Gold, R.~E., {et~al.} 2006, The Astrophysical Journal,
  649, 470

\bibitem[{Doschek \& Warren(2019)}]{Doschek2019}
Doschek, G.~A. \& Warren, H.~P. 2019, The Astrophysical Journal, 884, 158

\bibitem[{Du(2012)}]{Du2012}
Du, Z. 2012, Solar Physics, 278, 203, arXiv: 1112.5560 ISBN: 1573-093X

\bibitem[{D’Amicis {et~al.}(2021{\natexlab{a}})D’Amicis, Alielden, Perrone,
  Bruno, Telloni, Raines, Lepri, \& Zhao}]{DAmicis2021}
D’Amicis, R., Alielden, K., Perrone, D., {et~al.} 2021{\natexlab{a}},
  Astronomy \& Astrophysics, 654, A111

\bibitem[{D’Amicis {et~al.}(2021{\natexlab{b}})D’Amicis, Perrone, Bruno, \&
  Velli}]{DAmicis2021a}
D’Amicis, R., Perrone, D., Bruno, R., \& Velli, M. 2021{\natexlab{b}},
  Journal of Geophysical Research: Space Physics, 126

\bibitem[{Feldman \& Laming(2000)}]{Feldman2000}
Feldman, U. \& Laming, J.~M. 2000, Physica Scripta, 61, 222

\bibitem[{Feldman {et~al.}(1978)Feldman, Asbridge, Bame, \&
  Gosling}]{Feldman1978}
Feldman, W.~C., Asbridge, J.~R., Bame, S.~J., \& Gosling, J.~T. 1978, Journal
  of Geophysical Research, 83, 2177

\bibitem[{Fisk {et~al.}(1999)Fisk, Zurbuchen, \& Schwadron}]{Fisk1999}
Fisk, L.~A., Zurbuchen, T.~H., \& Schwadron, N.~A. 1999, The Astrophysical
  Journal, 521, 868

\bibitem[{Fu {et~al.}(2015)Fu, Li, Li, Huang, Mou, Jiao, \& Xia}]{Fu2015}
Fu, H., Li, B., Li, X., {et~al.} 2015, Solar Physics, 290, 1399

\bibitem[{Fu {et~al.}(2018)Fu, Madjarska, Li, Xia, \& Huang}]{Fu2018}
Fu, H., Madjarska, M.~S., Li, B., Xia, L., \& Huang, Z. 2018, Monthly Notices
  of the Royal Astronomical Society, 478, 1884

\bibitem[{Fu {et~al.}(2017)Fu, Madjarska, Xia, Li, Huang, \& Wangguan}]{Fu2017}
Fu, H., Madjarska, M.~S., Xia, L., {et~al.} 2017, The Astrophysical Journal,
  836, 169

\bibitem[{Geiss(1982)}]{Geiss1982a}
Geiss, J. 1982, Space Science Reviews, 33, 201, iSBN: 9783540773405

\bibitem[{Geiss {et~al.}(1995{\natexlab{a}})Geiss, Gloeckler, \& von
  Steiger}]{Geiss1995b}
Geiss, J., Gloeckler, G., \& von Steiger, R. 1995{\natexlab{a}}, Space Science
  Reviews, 72, 49

\bibitem[{Geiss {et~al.}(1995{\natexlab{b}})Geiss, Gloeckler, Von~Steiger,
  Balsiger, Fisk, Galvin, Ipavich, Livi, McKenzie, Ogilvie, Et, \&
  Wilken}]{Geiss1995}
Geiss, J., Gloeckler, G., Von~Steiger, R., {et~al.} 1995{\natexlab{b}},
  Science, 268, 1033, publisher: Physikalisches Institut, University of Bern,
  Switzerland. tex.ids= Geiss1995a

\bibitem[{Gloeckler {et~al.}(1998)Gloeckler, Cain, Ipavich, Tums, Bedini, Fisk,
  Zurbuchen, Bochsler, Fischer, Wimmer-Schweingruber, Geiss, Kallenbach, \&
  Kallenback}]{ACE:SWICS}
Gloeckler, G., Cain, J., Ipavich, F.~M., {et~al.} 1998, Space Sci. Rev., 86,
  497, publisher: Kluwer Academic Publishers

\bibitem[{Hansteen \& Velli(2012)}]{Hansteen2012}
Hansteen, V.~H. \& Velli, M. 2012, Space Science Reviews, 172, 89

\bibitem[{Hathaway(2015)}]{Hathaway2015}
Hathaway, D.~H. 2015, Living Reviews in Solar Physics, 12, iSBN: 1614-4961

\bibitem[{Hewins {et~al.}(2020)Hewins, Gibson, Webb, McFadden, Kuchar, Emery,
  \& McIntosh}]{Hewins2020}
Hewins, I.~M., Gibson, S.~E., Webb, D.~F., {et~al.} 2020, Solar Physics, 295,
  publisher: Springer Nature B.V.

\bibitem[{Hirshberg(1973)}]{Hirshberg1973}
Hirshberg, J. 1973, Reviews of Geophysics, 11, 115

\bibitem[{Holzer \& Leer(1980)}]{Holzer1980a}
Holzer, T.~E. \& Leer, E. 1980, Journal of Geophysical Research: Space Physics,
  85, 4665

\bibitem[{Holzer \& Leer(1981)}]{Holzer1981}
Holzer, T.~E. \& Leer, E. 1981, in Solar {Wind} 4 (Burhausen, Germany: Max
  Planck Institut für Aeronomie and Max Planck Institut für
  exraterrestriesche Physik), 28--41, conference Name: Solar Wind 4 Pages: 28
  ADS Bibcode: 1981sowi.conf...28H

\bibitem[{Johnson {et~al.}(2024)Johnson, Rivera, Niembro, Paulson, Badman,
  Stevens, Dieguez, Case, Bale, \& Kasper}]{Johnson2024}
Johnson, M., Rivera, Y.~J., Niembro, T., {et~al.} 2024, The Astrophysical
  Journal, 964, 81

\bibitem[{Johnstone {et~al.}(2015)Johnstone, Güdel, Lüftinger, Toth, \&
  Brott}]{Johnstone2015}
Johnstone, C.~P., Güdel, M., Lüftinger, T., Toth, G., \& Brott, I. 2015,
  Astronomy \& Astrophysics, 577, A27

\bibitem[{Kasper(2002)}]{KasperThesis}
Kasper, J.~C. 2002, PhD thesis, Massachusetts Institute of Technology

\bibitem[{Kasper {et~al.}(2021)Kasper, Klein, Lichko, Huang, Chen, Badman,
  Bonnell, Whittlesey, Livi, Larson, Pulupa, Rahmati, Stansby, Korreck,
  Stevens, Case, Bale, Maksimovic, Moncuquet, Goetz, Halekas, Malaspina,
  Raouafi, Szabo, MacDowall, Velli, Dudok De~Wit, \& Zank}]{PSP:SWEAP:Ra}
Kasper, J.~C., Klein, K.~G., Lichko, E., {et~al.} 2021, Physical Review
  Letters, 127, 255101

\bibitem[{Kasper {et~al.}(2017)Kasper, Klein, Weber, Maksimovic, Zaslavsky,
  Bale, Maruca, Stevens, \& Case}]{Kasper2017}
Kasper, J.~C., Klein, K.~G., Weber, T., {et~al.} 2017, The Astrophysical
  Journal, 849, 126

\bibitem[{Kasper {et~al.}(2008)Kasper, Lazarus, \& Gary}]{Kasper2008}
Kasper, J.~C., Lazarus, A.~J., \& Gary, S.~P. 2008, Physical Review Letters,
  101, 261103

\bibitem[{Kasper {et~al.}(2006)Kasper, Lazarus, Steinberg, Ogilvie, \&
  Szabo}]{Wind:SWE:bimax}
Kasper, J.~C., Lazarus, A.~J., Steinberg, J.~T., Ogilvie, K.~W., \& Szabo, A.
  2006, Journal of Geophysical Research, 111, A03105

\bibitem[{Kasper {et~al.}(2007)Kasper, Stevens, Lazarus, Steinberg, \&
  Ogilvie}]{Kasper:Ahe}
Kasper, J.~C., Stevens, M., Lazarus, A.~J., Steinberg, J.~T., \& Ogilvie, K.~W.
  2007, The Astrophysical Journal, 660, 901

\bibitem[{Kasper {et~al.}(2012)Kasper, Stevens, Korreck, Maruca, Kiefer,
  Schwadron, \& Lepri}]{Kasper:Ahe:Qstate}
Kasper, J.~C., Stevens, M.~L., Korreck, K.~E., {et~al.} 2012, The Astrophysical
  Journal, 745, 162

\bibitem[{Klein {et~al.}(2018)Klein, Alterman, Stevens, Vech, \&
  Kasper}]{Klein2018}
Klein, K.~G., Alterman, B.~L., Stevens, M., Vech, D., \& Kasper, J.~C. 2018,
  Physical Review Letters, 120, 205102, publisher: American Physical Society

\bibitem[{Laming(2004)}]{Laming2004}
Laming, J.~M. 2004, The Astrophysical Journal, 614, 1063

\bibitem[{Laming(2009)}]{Laming2009}
Laming, J.~M. 2009, The Astrophysical Journal, 695, 954

\bibitem[{Laming(2012)}]{Laming2012}
Laming, J.~M. 2012, The Astrophysical Journal, 744, 115, arXiv:
  astro-ph.SR/1110.4357

\bibitem[{Laming(2015)}]{LR:FIP}
Laming, J.~M. 2015, Living Reviews in Solar Physics, 12, arXiv: 1504.08325
  ISBN: 2367-3648

\bibitem[{Leer \& Holzer(1980)}]{Leer1980}
Leer, E. \& Holzer, T.~E. 1980, Journal of Geophysical Research: Space Physics,
  85, 4681

\bibitem[{Lepri {et~al.}(2013)Lepri, Landi, \& Zurbuchen}]{ACE:SWICS:AUX}
Lepri, S.~T., Landi, E., \& Zurbuchen, T.~H. 2013, The Astrophysical Journal,
  768, 94

\bibitem[{Lepri \& Rivera(2021)}]{Lepri2021}
Lepri, S.~T. \& Rivera, Y.~J. 2021, The Astrophysical Journal, 912, 51,
  publisher: IOP Publishing tex.ids= Lepri2021a, Lepri2021b

\bibitem[{Livi {et~al.}(2003)Livi, Möbius, Haggerty, Witte, \&
  Wurz}]{Livi2003}
Livi, S.~A., Möbius, E., Haggerty, D., Witte, M., \& Wurz, P. 2003, AIP
  Conference Proceedings, 679, 850, iSBN: 0735401489

\bibitem[{Marsch(2006)}]{Marsch2006b}
Marsch, E. 2006, Advances in Space Research, 38, 921

\bibitem[{Martinović {et~al.}(2021)Martinović, Klein, Durovcova, \&
  Alterman}]{Martinovic2021a}
Martinović, M., Klein, K.~G., Durovcova, T., \& Alterman, B.~L. 2021, 1,
  arXiv: 2110.07772

\bibitem[{Martinović {et~al.}(2020)Martinović, Klein, Kasper, Case, Korreck,
  Larson, Livi, Stevens, Whittlesey, Chandran, Alterman, Huang, Chen, Bale,
  Pulupa, Malaspina, Bonnell, Harvey, Goetz, Dudok~de Wit, \&
  MacDowall}]{Martinovic2019}
Martinović, M., Klein, K.~G., Kasper, J.~C., {et~al.} 2020, The Astrophysical
  Journal Supplement Series, 246, 30, arXiv: 1912.02653

\bibitem[{Maruca \& Kasper(2013)}]{Wind:SWE:dvapbimax}
Maruca, B.~A. \& Kasper, J.~C. 2013, Advances in Space Research, 52, 723,
  publisher: COSPAR

\bibitem[{Mason {et~al.}(2012)Mason, Desai, \& Li}]{Mason2012b}
Mason, G.~M., Desai, M., \& Li, G. 2012, Astrophysical Journal Letters, 748,
  2010

\bibitem[{McComas {et~al.}(2008)McComas, Ebert, Elliott, Goldstein, Gosling,
  Schwadron, \& Skoug}]{McComas2008}
McComas, D.~J., Ebert, R.~W., Elliott, H.~A., {et~al.} 2008, Geophysical
  Research Letters, 35, L18103

\bibitem[{McIntosh {et~al.}(2011)McIntosh, Kiefer, Leamon, Kasper, \&
  Stevens}]{McIntosh:Ahe}
McIntosh, S.~W., Kiefer, K.~K., Leamon, R.~J., Kasper, J.~C., \& Stevens, M.
  2011, Astrophysical Journal Letters, 740, 1, arXiv: 1109.1408

\bibitem[{McIntosh {et~al.}(2015)McIntosh, Leamon, Krista, Title, Hudson,
  Riley, Harder, Kopp, Snow, Woods, Kasper, Stevens, \& Ulrich}]{McIntosh2015a}
McIntosh, S.~W., Leamon, R.~J., Krista, L.~D., {et~al.} 2015, Nature
  Communications, 6, 1, publisher: Nature Publishing Group ISBN: 2041-1723
  (Electronic){\textbackslash}r2041-1723 (Linking)

\bibitem[{Meyer-Vernet(2007)}]{Meyer-Vernet2007a}
Meyer-Vernet, N. 2007, Basics of the {Solar} {Wind}, 1st edn. (Cambridge
  University Press)

\bibitem[{Mihailescu {et~al.}(2023)Mihailescu, Brooks, Laming, Baker, Green,
  James, Long, Van Driel-Gesztelyi, \& Stangalini}]{Mihailescu2023}
Mihailescu, T., Brooks, D.~H., Laming, J.~M., {et~al.} 2023, The Astrophysical
  Journal, 959, 72

\bibitem[{Nicolaou {et~al.}(2014)Nicolaou, Livadiotis, \&
  Moussas}]{Nicolaou2014}
Nicolaou, G., Livadiotis, G., \& Moussas, X. 2014, Solar Physics, 289, 1371

\bibitem[{Ogilvie {et~al.}(1995)Ogilvie, Chornay, Fritzenreiter, Hunsaker,
  Keller, Lobell, Miller, Scudder, Sittler, Torbert, Bodet, Needell, Lazarus,
  Steinberg, Tappan, Mavretic, \& Gergin}]{Wind:SWE}
Ogilvie, K.~W., Chornay, D.~J., Fritzenreiter, R.~J., {et~al.} 1995, Space
  Science Reviews, 71, 55

\bibitem[{Parker(1958)}]{Parker1958a}
Parker, E.~N. 1958, The Astrophysical Journal, 128, 664, iSBN: 9780874216561

\bibitem[{Phillips {et~al.}(1994)Phillips, Balogh, Bame, Goldstein, Gosling,
  Hoeksema, McComas, Neugebauer, Sheeley, \& Wang}]{Phillips1994}
Phillips, J.~L., Balogh, A., Bame, S.~J., {et~al.} 1994, Geophysical Research
  Letters, 21, 1105

\bibitem[{Pilleri {et~al.}(2015)Pilleri, Reisenfeld, Zurbuchen, Lepri, Shearer,
  Gilbert, Steiger, \& Wiens}]{Pilleri2015}
Pilleri, P., Reisenfeld, D.~B., Zurbuchen, T.~H., {et~al.} 2015, The
  Astrophysical Journal, 812, 1

\bibitem[{Pottasch(1963)}]{Pottasch1963}
Pottasch, S.~R. 1963, The Astrophysical Journal, 137, 945

\bibitem[{Raouafi {et~al.}(2023)Raouafi, Stenborg, Seaton, Wang, Wang,
  DeForest, Bale, Drake, Uritsky, Karpen, DeVore, Sterling, Horbury, Harra,
  Bourouaine, Kasper, Kumar, Phan, \& Velli}]{Raouafi2023}
Raouafi, N.~E., Stenborg, G., Seaton, D.~B., {et~al.} 2023, The Astrophysical
  Journal, 945, 28

\bibitem[{Raymond {et~al.}(1997)Raymond, Kohl, Noci, Antonucci, Tondello,
  Huber, Gardner, Nicolosi, Fineschi, Romoli, Spadaro, Siegmund, Benna,
  Ciaravella, Cranmer, Giordano, Karovska, Martin, Michels, Modigliani,
  Naletto, Panasyuk, Pernechele, Poletto, Smith, Suleiman, \&
  Strachan}]{Raymond1997}
Raymond, J.~C., Kohl, J.~L., Noci, G., {et~al.} 1997, in The {First} {Results}
  from {SOHO}, ed. B.~Fleck \& Z.~Švestka (Dordrecht: Springer Netherlands),
  645--665

\bibitem[{Reisenfeld {et~al.}(2013)Reisenfeld, Wiens, Barraclough, Steinberg,
  Neugebauer, Raines, \& Zurbuchen}]{Reisenfeld2013}
Reisenfeld, D.~B., Wiens, R.~C., Barraclough, B.~L., {et~al.} 2013, Space
  Science Reviews, 175, 125

\bibitem[{Renaud \& Victoria-Feser(2010)}]{Renaud2010}
Renaud, O. \& Victoria-Feser, M.-P. 2010, Journal of Statistical Planning and
  Inference, 140, 1852

\bibitem[{Richardson \& Cane(2010)}]{Richardson2010}
Richardson, I.~G. \& Cane, H.~V. 2010, Solar Physics, 264, 189

\bibitem[{Rivera {et~al.}(2024{\natexlab{a}})Rivera, Badman, Stevens, Raines,
  Owen, Paulson, Niembro, Livi, Lepri, Landi, Halekas, Ervin, Dewey, Coburn,
  Bale, \& Alterman}]{Rivera2024a}
Rivera, Y.~J., Badman, S.~T., Stevens, M.~L., {et~al.} 2024{\natexlab{a}}, The
  Astrophysical Journal, 974, 198

\bibitem[{Rivera {et~al.}(2024{\natexlab{b}})Rivera, Badman, Stevens, Verniero,
  Stawarz, Shi, Raines, Paulson, Owen, Niembro, Louarn, Livi, Lepri, Kasper,
  Horbury, Halekas, Dewey, De~Marco, \& Bale}]{Rivera2024}
Rivera, Y.~J., Badman, S.~T., Stevens, M.~L., {et~al.} 2024{\natexlab{b}},
  Science, 385, 962

\bibitem[{Rivera {et~al.}(2022{\natexlab{a}})Rivera, Higginson, Lepri, Viall,
  Alterman, Landi, Spitzer, Raines, Cranmer, Laming, Mason, Wallace, Raymond,
  Lynch, Gilly, Chen, \& Dewey}]{Rivera2022a}
Rivera, Y.~J., Higginson, A., Lepri, S.~T., {et~al.} 2022{\natexlab{a}},
  Frontiers in Astronomy and Space Sciences, 9, 1056347

\bibitem[{Rivera {et~al.}(2020)Rivera, Landi, Lepri, \& Gilbert}]{Rivera2020}
Rivera, Y.~J., Landi, E., Lepri, S.~T., \& Gilbert, J.~A. 2020, The
  Astrophysical Journal, 899, 11

\bibitem[{Rivera {et~al.}(2021)Rivera, Lepri, Raymond, Reeves, Stevens, \&
  Zhao}]{Rivera2021}
Rivera, Y.~J., Lepri, S.~T., Raymond, J.~C., {et~al.} 2021, The Astrophysical
  Journal, 921, 93

\bibitem[{Rivera {et~al.}(2022{\natexlab{b}})Rivera, Raymond, Landi, Lepri,
  Reeves, Stevens, \& Alterman}]{Rivera2022}
Rivera, Y.~J., Raymond, J.~C., Landi, E., {et~al.} 2022{\natexlab{b}}, The
  Astrophysical Journal, 936, 83

\bibitem[{Schwadron {et~al.}(1999)Schwadron, Fisk, \&
  Zurbuchen}]{Schwadron1999}
Schwadron, N.~A., Fisk, L.~A., \& Zurbuchen, T.~H. 1999, The Astrophysical
  Journal, 521, 859

\bibitem[{Schwenn(2006)}]{Schwenn2006}
Schwenn, R. 2006, Space Science Reviews, 124, 51

\bibitem[{Shearer {et~al.}(2014)Shearer, von Steiger, Raines, Lepri, Thomas,
  Gilbert, Landi, \& Zurbuchen}]{ACE:SWICS:MLE}
Shearer, P., von Steiger, R., Raines, J.~M., {et~al.} 2014, The Astrophysical
  Journal, 789, 60

\bibitem[{Song {et~al.}(2020)Song, Zhang, Cheng, Li, Hu, Li, Chen, Zheng, \&
  Chen}]{Song2020}
Song, H.~Q., Zhang, J., Cheng, X., {et~al.} 2020, The Astrophysical Journal,
  901, L21, arXiv: 2009.05212

\bibitem[{Stakhiv {et~al.}(2015)Stakhiv, Landi, Lepri, Oran, \&
  Zurbuchen}]{Stakhiv2015}
Stakhiv, M.~O., Landi, E., Lepri, S.~T., Oran, R., \& Zurbuchen, T.~H. 2015,
  The Astrophysical Journal, 801, 100

\bibitem[{Stakhiv {et~al.}(2016)Stakhiv, Lepri, Landi, Tracy, \&
  Zurbuchen}]{Stakhiv2016}
Stakhiv, M.~O., Lepri, S.~T., Landi, E., Tracy, P.~J., \& Zurbuchen, T.~H.
  2016, The Astrophysical Journal, 829, 117, publisher: IOP Publishing ISBN:
  0769518745

\bibitem[{Stone {et~al.}(1998)Stone, Frandsen, Mewaldt, Christian, Margolies,
  Ormes, \& Snow}]{ACE}
Stone, E.~C., Frandsen, A.~M., Mewaldt, R.~A., {et~al.} 1998, Space Science
  Reviews, 86, 1, publisher: Kluwer Academic Publishers ISBN:
  10.1023/A:1005082526237

\bibitem[{Subramanian {et~al.}(2010)Subramanian, Madjarska, \&
  Doyle}]{Subramanian2010}
Subramanian, S., Madjarska, M.~S., \& Doyle, J.~G. 2010, Astronomy and
  Astrophysics, 516, A50

\bibitem[{Tlatov {et~al.}(2014)Tlatov, Tavastsherna, \& Vasil'eva}]{Tlatov2014}
Tlatov, A., Tavastsherna, K., \& Vasil'eva, V. 2014, Solar Physics, 289, 1349

\bibitem[{Tracy {et~al.}(2016)Tracy, Kasper, Raines, Shearer, Gilbert, \&
  Zurbuchen}]{Tracy2016}
Tracy, P.~J., Kasper, J.~C., Raines, J.~M., {et~al.} 2016, Physical Review
  Letters, 255101, 255101

\bibitem[{Tu \& Marsch(1995)}]{Tu1995}
Tu, C.~Y. \& Marsch, E. 1995, Space Science Reviews, 73, 1, iSBN: 0038-6308

\bibitem[{Verscharen {et~al.}(2019)Verscharen, Klein, \& Maruca}]{LR:MSSW}
Verscharen, D., Klein, K.~G., \& Maruca, B.~A. 2019, The multi-scale nature of
  the solar wind, Vol.~16 (Springer International Publishing), arXiv:
  1902.03448 Publication Title: Living Reviews in Solar Physics Issue: 1 ISSN:
  16144961

\bibitem[{Viall \& Borovsky(2020)}]{Viall:9Q}
Viall, N.~M. \& Borovsky, J. 2020, Journal of Geophysical Research: Space
  Physics, 125, 1

\bibitem[{von Steiger {et~al.}(1997)von Steiger, Geiss, \&
  Gloeckler}]{vonSteiger1997}
von Steiger, R., Geiss, J., \& Gloeckler, G. 1997, in Cosmic {Winds} and the
  {Heliosphere}, Space {Science} {Series}, 581--611, conference Name: Cosmic
  Winds and the Heliosphere Pages: 581 ADS Bibcode: 1997cwh..conf..581V

\bibitem[{von Steiger {et~al.}(2000)von Steiger, Schwadron, Fisk, Geiss,
  Gloeckler, Hefti, Wilken, Wimmer-Schweingruber, \&
  Zurbuchen}]{vonSteiger2000}
von Steiger, R., Schwadron, N.~A., Fisk, L.~A., {et~al.} 2000, Journal of
  Geophysical Research: Space Physics, 105, 27217

\bibitem[{Von~Steiger \& Zurbuchen(2016)}]{VonSteiger2016}
Von~Steiger, R. \& Zurbuchen, T.~H. 2016, The Astrophysical Journal, 816, 13

\bibitem[{Wang \& Sheeley(2002)}]{Wang2002}
Wang, Y.~M. \& Sheeley, N.~R. 2002, Journal of Geophysical Research: Space
  Physics, 107, 1

\bibitem[{Weberg(2015)}]{Weberg2015}
Weberg, M. 2015, Spatial and {Temporal} {Coordinate} {Systems} in {Space}
  {Physics}, Tech. rep.

\bibitem[{Weberg {et~al.}(2012)Weberg, Zurbuchen, \& Lepri}]{Weberg2012}
Weberg, M., Zurbuchen, T.~H., \& Lepri, S.~T. 2012, Astrophysical Journal, 760

\bibitem[{Widing \& Feldman(2001)}]{Widing2001}
Widing, K.~G. \& Feldman, U. 2001, The Astrophysical Journal, 555, 426

\bibitem[{Wurz {et~al.}(2000)Wurz, Bochsler, \& Lee}]{Wurz2000}
Wurz, P., Bochsler, P., \& Lee, M.~A. 2000, Journal of Geophysical Research:
  Space Physics, 105, 27239

\bibitem[{Xu \& Borovsky(2015)}]{Xu2014}
Xu, F. \& Borovsky, J. 2015, Journal of Geophysical Research: Space Physics,
  120, 70

\bibitem[{{Yogesh} {et~al.}(2021){Yogesh}, Chakrabarty, \&
  Srivastava}]{Yogesh2021}
{Yogesh}, Chakrabarty, D., \& Srivastava, N. 2021, Monthly Notices of the Royal
  Astronomical Society: Letters, 503, L17, publisher: Oxford University Press

\bibitem[{{Yogesh} {et~al.}(2023){Yogesh}, Chakrabarty, \&
  Srivastava}]{Yogesh2023}
{Yogesh}, Chakrabarty, D., \& Srivastava, N. 2023, Monthly Notices of the Royal
  Astronomical Society, 526, L13

\bibitem[{Zerbo \& Richardson(2015)}]{Zerbo2015}
Zerbo, J.-L. \& Richardson, J.~D. 2015, Journal of Geophysical Research: Space
  Physics, 120, 10,250

\bibitem[{Zhao {et~al.}(2022)Zhao, Landi, Lepri, \& Carpenter}]{Zhao2022}
Zhao, L., Landi, E., Lepri, S.~T., \& Carpenter, D. 2022, Universe, 8, 393

\bibitem[{Zhao {et~al.}(2017{\natexlab{a}})Zhao, Landi, Lepri, Gilbert,
  Zurbuchen, Fisk, \& Raines}]{Zhao:InSituComposition:Sources}
Zhao, L., Landi, E., Lepri, S.~T., {et~al.} 2017{\natexlab{a}}, The
  Astrophysical Journal, 846, 135, publisher: IOP Publishing

\bibitem[{Zhao {et~al.}(2017{\natexlab{b}})Zhao, Zhang, \& Rassoul}]{Zhao2017a}
Zhao, L., Zhang, M., \& Rassoul, H.~K. 2017{\natexlab{b}}, The Astrophysical
  Journal, 836, 31, publisher: IOP Publishing

\bibitem[{Zurbuchen \& Richardson(2006)}]{Zurbuchen2006}
Zurbuchen, T.~H. \& Richardson, I.~G. 2006, Space Science Reviews, 123, 31

\bibitem[{Zurbuchen {et~al.}(2016)Zurbuchen, Weberg, Von~Steiger, Mewaldt,
  Lepri, \& Antiochos}]{Zurbuchen2016}
Zurbuchen, T.~H., Weberg, M., Von~Steiger, R., {et~al.} 2016, The Astrophysical
  Journal, 826, 10

\end{thebibliography}


\begin{thebibliography}{}
\expandafter\ifx\csname natexlab\endcsname\relax\def\natexlab#1{#1}\fi
\providecommand{\url}[1]{\href{#1}{#1}}
\providecommand{\dodoi}[1]{doi:~\href{http://doi.org/#1}{\nolinkurl{#1}}}
\providecommand{\doeprint}[1]{\href{http://ascl.net/#1}{\nolinkurl{http://ascl.net/#1}}}
\providecommand{\doarXiv}[1]{\href{https://arxiv.org/abs/#1}{\nolinkurl{https://arxiv.org/abs/#1}}}

\bibitem[{Abbo {et~al.}(2016)Abbo, Ofman, Antiochos, Hansteen, Harra, Ko,
  Lapenta, Li, Riley, Strachan, von Steiger, \& Wang}]{Abbo2016}
Abbo, L., Ofman, L., Antiochos, S.~K., {et~al.} 2016, Space Science Reviews,
  201, 55, \dodoi{10.1007/s11214-016-0264-1}

\bibitem[{Aellig {et~al.}(2001)Aellig, Lazarus, \& Steinberg}]{Aellig2001}
Aellig, M.~R., Lazarus, A.~J., \& Steinberg, J.~T. 2001, Geophysical Research
  Letters, 28, 2767, \dodoi{10.1029/2000GL012771}

\bibitem[{Alterman {et~al.}(2023)Alterman, Desai, Dayeh, Mason, \&
  Ho}]{STQT:selection-abundances}
Alterman, B.~L., Desai, M.~I., Dayeh, M.~A., Mason, G.~M., \& Ho, G. 2023, The
  Astrophysical Journal, 952, 42, \dodoi{10.3847/1538-4357/acd24a}

\bibitem[{Alterman \& Kasper(2019)}]{Alterman2019}
Alterman, B.~L., \& Kasper, J.~C. 2019, The Astrophysical Journal, 879, L6,
  \dodoi{10.3847/2041-8213/ab2391}

\bibitem[{Alterman {et~al.}(2021)Alterman, Kasper, Leamon, \&
  McIntosh}]{Alterman2021}
Alterman, B.~L., Kasper, J.~C., Leamon, R.~J., \& McIntosh, S.~W. 2021, Solar
  Physics, 296, 67, \dodoi{10.1007/s11207-021-01801-9}

\bibitem[{Alterman {et~al.}(2018)Alterman, Kasper, Stevens, \&
  Koval}]{Alterman2018}
Alterman, B.~L., Kasper, J.~C., Stevens, M., \& Koval, A. 2018, The
  Astrophysical Journal, 864, 112, \dodoi{10.3847/1538-4357/aad23f}

\bibitem[{Antiochos {et~al.}(2011)Antiochos, Mikic, Titov, Lionello, \&
  Linker}]{Antiochos2011}
Antiochos, S.~K., Mikic, Z., Titov, V.~S., Lionello, R., \& Linker, J.~A. 2011,
  The Astrophysical Journal, 112, \dodoi{10.1088/0004-637X/731/2/112}

\bibitem[{Antonucci {et~al.}(2005)Antonucci, Abbo, \& Dodero}]{Antonucci2005}
Antonucci, E., Abbo, L., \& Dodero, M.~A. 2005, Astronomy \& Astrophysics, 435,
  699, \dodoi{10.1051/0004-6361:20047126}

\bibitem[{Asplund {et~al.}(2021)Asplund, Amarsi, \& Grevesse}]{Asplund2021}
Asplund, M., Amarsi, A.~M., \& Grevesse, N. 2021, Astronomy \& Astrophysics,
  653, A141, \dodoi{10.1051/0004-6361/202140445}

\bibitem[{Bodmer \& Bochsler(1998)}]{Bodmer1998a}
Bodmer, R., \& Bochsler, P. 1998, Physics and Chemistry of the Earth, 23, 683,
  \dodoi{10.1016/S0079-1946(98)00111-6}

\bibitem[{Borovsky(2016)}]{Borovsky2016}
Borovsky, J. 2016, Journal of Geophysical Research A: Space Physics, 121, 5055,
  \dodoi{10.1002/2016JA022686}

\bibitem[{Brooks {et~al.}(2015)Brooks, Ugarte-Urra, \& Warren}]{Brooks2015}
Brooks, D.~H., Ugarte-Urra, I., \& Warren, H.~P. 2015, Nature Communications,
  6, \dodoi{10.1038/ncomms6947}

\bibitem[{Bruno \& Carbone(2013)}]{LR:turbulence}
Bruno, R., \& Carbone, V. 2013, Living Reviews in Solar Physics, 10, 1,
  \dodoi{10.12942/lrsp-2013-2}

\bibitem[{Burnett {et~al.}(2003)Burnett, Barraclough, Bennett, Neugebauer,
  Oldham, Sasaki, Sevilla, Smith, Stansbery, Sweetnam, \& Wiens}]{Genesis}
Burnett, D.~S., Barraclough, B.~L., Bennett, R., {et~al.} 2003, Space Science
  Reviews, 105, 509, \dodoi{10.1023/A:1024425810605}

\bibitem[{Crooker {et~al.}(2012)Crooker, Antiochos, Zhao, \&
  Neugebauer}]{Crooker2012}
Crooker, N.~U., Antiochos, S.~K., Zhao, X., \& Neugebauer, M. 2012, Journal of
  Geophysical Research: Space Physics, 117, n/a, \dodoi{10.1029/2011JA017236}

\bibitem[{D'Amicis \& Bruno(2015)}]{DAmicis2015}
D'Amicis, R., \& Bruno, R. 2015, Astrophysical Journal, 805, 1,
  \dodoi{10.1088/0004-637X/805/1/84}

\bibitem[{Desai {et~al.}(2006)Desai, Mason, Gold, Krimigis, Cohen, Mewaldt,
  Mazur, \& Dwyer}]{Desai2006b}
Desai, M., Mason, G., Gold, R.~E., {et~al.} 2006, The Astrophysical Journal,
  649, 470, \dodoi{10.1086/505649}

\bibitem[{Du(2012)}]{Du2012}
Du, Z. 2012, Solar Physics, 278, 203, \dodoi{10.1007/s11207-011-9925-0}

\bibitem[{D’Amicis {et~al.}(2021{\natexlab{a}})D’Amicis, Alielden, Perrone,
  Bruno, Telloni, Raines, Lepri, \& Zhao}]{DAmicis2021}
D’Amicis, R., Alielden, K., Perrone, D., {et~al.} 2021{\natexlab{a}},
  Astronomy \& Astrophysics, 654, A111, \dodoi{10.1051/0004-6361/202140600}

\bibitem[{D’Amicis {et~al.}(2021{\natexlab{b}})D’Amicis, Perrone, Bruno, \&
  Velli}]{DAmicis2021a}
D’Amicis, R., Perrone, D., Bruno, R., \& Velli, M. 2021{\natexlab{b}},
  Journal of Geophysical Research: Space Physics, 126,
  \dodoi{10.1029/2020JA028996}

\bibitem[{Feldman \& Laming(2000)}]{Feldman2000}
Feldman, U., \& Laming, J.~M. 2000, Physica Scripta, 61, 222,
  \dodoi{10.1238/Physica.Regular.061a00222}

\bibitem[{Feldman {et~al.}(1978)Feldman, Asbridge, Bame, \&
  Gosling}]{Feldman1978}
Feldman, W.~C., Asbridge, J.~R., Bame, S.~J., \& Gosling, J.~T. 1978, Journal
  of Geophysical Research, 83, 2177, \dodoi{10.1029/JA083iA05p02177}

\bibitem[{Fisk {et~al.}(1999)Fisk, Zurbuchen, \& Schwadron}]{Fisk1999}
Fisk, L.~A., Zurbuchen, T.~H., \& Schwadron, N.~A. 1999, The Astrophysical
  Journal, 521, 868, \dodoi{10.1086/307556}

\bibitem[{Fu {et~al.}(2015)Fu, Li, Li, Huang, Mou, Jiao, \& Xia}]{Fu2015}
Fu, H., Li, B., Li, X., {et~al.} 2015, Solar Physics, 290, 1399,
  \dodoi{10.1007/s11207-015-0689-9}

\bibitem[{Fu {et~al.}(2018)Fu, Madjarska, Li, Xia, \& Huang}]{Fu2018}
Fu, H., Madjarska, M.~S., Li, B., Xia, L., \& Huang, Z. 2018, Monthly Notices
  of the Royal Astronomical Society, 478, 1884, \dodoi{10.1093/mnras/sty1211}

\bibitem[{Fu {et~al.}(2017)Fu, Madjarska, Xia, Li, Huang, \& Wangguan}]{Fu2017}
Fu, H., Madjarska, M.~S., Xia, L., {et~al.} 2017, The Astrophysical Journal,
  836, 169, \dodoi{10.3847/1538-4357/aa5cba}

\bibitem[{Geiss {et~al.}(1995{\natexlab{a}})Geiss, Gloeckler, \& von
  Steiger}]{Geiss1995b}
Geiss, J., Gloeckler, G., \& von Steiger, R. 1995{\natexlab{a}}, Space Science
  Reviews, 72, 49

\bibitem[{Geiss {et~al.}(1995{\natexlab{b}})Geiss, Gloeckler, Von~Steiger,
  Balsiger, Fisk, Galvin, Ipavich, Livi, McKenzie, Ogilvie, Et, \&
  Wilken}]{Geiss1995}
Geiss, J., Gloeckler, G., Von~Steiger, R., {et~al.} 1995{\natexlab{b}},
  Science, 268, 1033, \dodoi{10.1126/science.7754380}

\bibitem[{Gloeckler {et~al.}(1998)Gloeckler, Cain, Ipavich, Tums, Bedini, Fisk,
  Zurbuchen, Bochsler, Fischer, Wimmer-Schweingruber, Geiss, Kallenbach, \&
  Kallenback}]{ACE:SWICS}
Gloeckler, G., Cain, J., Ipavich, F.~M., {et~al.} 1998, Space Sci. Rev., 86,
  497, \dodoi{10.1023/A:1005036131689}

\bibitem[{Hansteen \& Velli(2012)}]{Hansteen2012}
Hansteen, V.~H., \& Velli, M. 2012, Space Science Reviews, 172, 89,
  \dodoi{10.1007/s11214-012-9887-z}

\bibitem[{Hathaway(2015)}]{Hathaway2015}
Hathaway, D.~H. 2015, Living Reviews in Solar Physics, 12,
  \dodoi{10.1007/lrsp-2015-4}

\bibitem[{Hewins {et~al.}(2020)Hewins, Gibson, Webb, McFadden, Kuchar, Emery,
  \& McIntosh}]{Hewins2020}
Hewins, I.~M., Gibson, S.~E., Webb, D.~F., {et~al.} 2020, Solar Physics, 295,
  \dodoi{10.1007/s11207-020-01731-y}

\bibitem[{Hirshberg(1973)}]{Hirshberg1973}
Hirshberg, J. 1973, Reviews of Geophysics, 11, 115,
  \dodoi{10.1029/RG011i001p00115}

\bibitem[{Holzer \& Leer(1980)}]{Holzer1980a}
Holzer, T.~E., \& Leer, E. 1980, Journal of Geophysical Research: Space
  Physics, 85, 4665, \dodoi{10.1029/JA085iA09p04665}

\bibitem[{Holzer \& Leer(1981)}]{Holzer1981}
Holzer, T.~E., \& Leer, E. 1981 (Burhausen, Germany: Max Planck Institut für
  Aeronomie and Max Planck Institut für exraterrestriesche Physik), 28--41.
\newblock \url{https://ui.adsabs.harvard.edu/abs/1981sowi.conf...28H}

\bibitem[{Ipython {et~al.}(????)Ipython, Print, \& Gui}]{IPython}
Ipython, M., Print, I., \& Gui, I. ????, The following magic functions are
  currently available :

\bibitem[{Johnson {et~al.}(2024)Johnson, Rivera, Niembro, Paulson, Badman,
  Stevens, Dieguez, Case, Bale, \& Kasper}]{Johnson2024}
Johnson, M., Rivera, Y.~J., Niembro, T., {et~al.} 2024, The Astrophysical
  Journal, 964, 81, \dodoi{10.3847/1538-4357/ad2510}

\bibitem[{Johnstone {et~al.}(2015)Johnstone, Güdel, Lüftinger, Toth, \&
  Brott}]{Johnstone2015}
Johnstone, C.~P., Güdel, M., Lüftinger, T., Toth, G., \& Brott, I. 2015,
  Astronomy \& Astrophysics, 577, A27, \dodoi{10.1051/0004-6361/201425300}

\bibitem[{Kasper {et~al.}(2021)Kasper, Klein, Lichko, Huang, Chen, Badman,
  Bonnell, Whittlesey, Livi, Larson, Pulupa, Rahmati, Stansby, Korreck,
  Stevens, Case, Bale, Maksimovic, Moncuquet, Goetz, Halekas, Malaspina,
  Raouafi, Szabo, MacDowall, Velli, Dudok De~Wit, \& Zank}]{Kasper2021}
Kasper, J., Klein, K., Lichko, E., {et~al.} 2021, Physical Review Letters, 127,
  255101, \dodoi{10.1103/PhysRevLett.127.255101}

\bibitem[{Kasper(2002)}]{KasperThesis}
Kasper, J.~C. 2002, PhD thesis, Massachusetts Institute of Technology.
\newblock \url{http://dspace.mit.edu/handle/1721.1/29937}

\bibitem[{Kasper {et~al.}(2008)Kasper, Lazarus, \& Gary}]{Kasper2008}
Kasper, J.~C., Lazarus, A.~J., \& Gary, S.~P. 2008, Physical Review Letters,
  101, 261103, \dodoi{10.1103/PhysRevLett.101.261103}

\bibitem[{Kasper {et~al.}(2006)Kasper, Lazarus, Steinberg, Ogilvie, \&
  Szabo}]{Wind:SWE:bimax}
Kasper, J.~C., Lazarus, A.~J., Steinberg, J.~T., Ogilvie, K.~W., \& Szabo, A.
  2006, Journal of Geophysical Research, 111, A03105,
  \dodoi{10.1029/2005JA011442}

\bibitem[{Kasper {et~al.}(2007)Kasper, Stevens, Lazarus, Steinberg, \&
  Ogilvie}]{Kasper2007}
Kasper, J.~C., Stevens, M., Lazarus, A.~J., Steinberg, J.~T., \& Ogilvie, K.~W.
  2007, The Astrophysical Journal, 660, 901, \dodoi{10.1086/510842}

\bibitem[{Kasper {et~al.}(2012)Kasper, Stevens, Korreck, Maruca, Kiefer,
  Schwadron, \& Lepri}]{Kasper2012}
Kasper, J.~C., Stevens, M.~L., Korreck, K.~E., {et~al.} 2012, The Astrophysical
  Journal, 745, 162, \dodoi{10.1088/0004-637X/745/2/162}

\bibitem[{Kasper {et~al.}(2017)Kasper, Klein, Weber, Maksimovic, Zaslavsky,
  Bale, Maruca, Stevens, \& Case}]{Kasper2017}
Kasper, J.~C., Klein, K.~G., Weber, T., {et~al.} 2017, The Astrophysical
  Journal, 849, 126, \dodoi{10.3847/1538-4357/aa84b1}

\bibitem[{Klein {et~al.}(2018)Klein, Alterman, Stevens, Vech, \&
  Kasper}]{Klein2018}
Klein, K.~G., Alterman, B.~L., Stevens, M., Vech, D., \& Kasper, J.~C. 2018,
  Physical Review Letters, 120, 205102, \dodoi{10.1103/PhysRevLett.120.205102}

\bibitem[{Laming(2004)}]{Laming2004}
Laming, J.~M. 2004, The Astrophysical Journal, 614, 1063,
  \dodoi{10.1086/423780}

\bibitem[{Laming(2009)}]{Laming2009}
---. 2009, The Astrophysical Journal, 695, 954,
  \dodoi{10.1088/0004-637X/695/2/954}

\bibitem[{Laming(2012)}]{Laming2012}
---. 2012, The Astrophysical Journal, 744, 115,
  \dodoi{10.1088/0004-637X/744/2/115}

\bibitem[{Laming(2015)}]{LR:FIP}
---. 2015, Living Reviews in Solar Physics, 12, \dodoi{10.1007/lrsp-2015-2}

\bibitem[{Leer \& Holzer(1980)}]{Leer1980}
Leer, E., \& Holzer, T.~E. 1980, Journal of Geophysical Research: Space
  Physics, 85, 4681, \dodoi{10.1029/JA085iA09p04681}

\bibitem[{Lepri {et~al.}(2013)Lepri, Landi, \& Zurbuchen}]{ACE:SWICS:AUX}
Lepri, S.~T., Landi, E., \& Zurbuchen, T.~H. 2013, The Astrophysical Journal,
  768, 94, \dodoi{10.1088/0004-637X/768/1/94}

\bibitem[{Lepri \& Rivera(2021)}]{Lepri2021}
Lepri, S.~T., \& Rivera, Y.~J. 2021, The Astrophysical Journal, 912, 51,
  \dodoi{10.3847/1538-4357/abea9f}

\bibitem[{Livi {et~al.}(2003)Livi, Möbius, Haggerty, Witte, \&
  Wurz}]{Livi2003}
Livi, S.~A., Möbius, E., Haggerty, D., Witte, M., \& Wurz, P. 2003, AIP
  Conference Proceedings, 679, 850, \dodoi{10.1063/1.1618724}

\bibitem[{Marsch(2006)}]{Marsch2006b}
Marsch, E. 2006, Advances in Space Research, 38, 921,
  \dodoi{10.1016/j.asr.2005.07.029}

\bibitem[{Martinović {et~al.}(2021)Martinović, Klein, Durovcova, \&
  Alterman}]{Martinovic2021a}
Martinović, M., Klein, K.~G., Durovcova, T., \& Alterman, B.~L. 2021, 1.
\newblock \url{http://arxiv.org/abs/2110.07772}

\bibitem[{Martinović {et~al.}(2020)Martinović, Klein, Kasper, Case, Korreck,
  Larson, Livi, Stevens, Whittlesey, Chandran, Alterman, Huang, Chen, Bale,
  Pulupa, Malaspina, Bonnell, Harvey, Goetz, Dudok~de Wit, \&
  MacDowall}]{Martinovic2019}
Martinović, M., Klein, K.~G., Kasper, J.~C., {et~al.} 2020, The Astrophysical
  Journal Supplement Series, 246, 30, \dodoi{10.3847/1538-4365/ab527f}

\bibitem[{Maruca \& Kasper(2013)}]{Wind:SWE:dvapbimax}
Maruca, B.~A., \& Kasper, J.~C. 2013, Advances in Space Research, 52, 723,
  \dodoi{10.1016/j.asr.2013.04.006}

\bibitem[{Mason {et~al.}(2012)Mason, Desai, \& Li}]{Mason2012b}
Mason, G., Desai, M., \& Li, G. 2012, Astrophysical Journal Letters, 748, 2010,
  \dodoi{10.1088/2041-8205/748/2/L31}

\bibitem[{McComas {et~al.}(2008)McComas, Ebert, Elliott, Goldstein, Gosling,
  Schwadron, \& Skoug}]{McComas2008}
McComas, D.~J., Ebert, R.~W., Elliott, H.~A., {et~al.} 2008, Geophysical
  Research Letters, 35, L18103, \dodoi{10.1029/2008GL034896}

\bibitem[{McIntosh {et~al.}(2011)McIntosh, Kiefer, Leamon, Kasper, \&
  Stevens}]{McIntosh2011a}
McIntosh, S.~W., Kiefer, K.~K., Leamon, R.~J., Kasper, J.~C., \& Stevens, M.
  2011, Astrophysical Journal Letters, 740, 1,
  \dodoi{10.1088/2041-8205/740/1/L23}

\bibitem[{McIntosh {et~al.}(2015)McIntosh, Leamon, Krista, Title, Hudson,
  Riley, Harder, Kopp, Snow, Woods, Kasper, Stevens, \& Ulrich}]{McIntosh2015a}
McIntosh, S.~W., Leamon, R.~J., Krista, L.~D., {et~al.} 2015, Nature
  Communications, 6, 1, \dodoi{10.1038/ncomms7491}

\bibitem[{Nicolaou {et~al.}(2014)Nicolaou, Livadiotis, \&
  Moussas}]{Nicolaou2014}
Nicolaou, G., Livadiotis, G., \& Moussas, X. 2014, Solar Physics, 289, 1371,
  \dodoi{10.1007/s11207-013-0401-x}

\bibitem[{Ogilvie {et~al.}(1995)Ogilvie, Chornay, Fritzenreiter, Hunsaker,
  Keller, Lobell, Miller, Scudder, Sittler, Torbert, Bodet, Needell, Lazarus,
  Steinberg, Tappan, Mavretic, \& Gergin}]{Wind:SWE}
Ogilvie, K.~W., Chornay, D.~J., Fritzenreiter, R.~J., {et~al.} 1995, Space
  Science Reviews, 71, 55, \dodoi{10.1007/BF00751326}

\bibitem[{Parker(1958)}]{Parker1958a}
Parker, E.~N. 1958, The Astrophysical Journal, 128, 664, \dodoi{10.1086/146579}

\bibitem[{Phillips {et~al.}(1994)Phillips, Balogh, Bame, Goldstein, Gosling,
  Hoeksema, McComas, Neugebauer, Sheeley, \& Wang}]{Phillips1994}
Phillips, J.~L., Balogh, A., Bame, S.~J., {et~al.} 1994, Geophysical Research
  Letters, 21, 1105, \dodoi{10.1029/94GL01065}

\bibitem[{Pilleri {et~al.}(2015)Pilleri, Reisenfeld, Zurbuchen, Lepri, Shearer,
  Gilbert, Steiger, \& Wiens}]{Pilleri2015}
Pilleri, P., Reisenfeld, D.~B., Zurbuchen, T.~H., {et~al.} 2015, The
  Astrophysical Journal, 812, 1, \dodoi{10.1088/0004-637X/812/1/1}

\bibitem[{Pottasch(1963)}]{Pottasch1963}
Pottasch, S.~R. 1963, The Astrophysical Journal, 137, 945,
  \dodoi{10.1086/147569}

\bibitem[{Reisenfeld {et~al.}(2013)Reisenfeld, Wiens, Barraclough, Steinberg,
  Neugebauer, Raines, \& Zurbuchen}]{Reisenfeld2013}
Reisenfeld, D.~B., Wiens, R.~C., Barraclough, B.~L., {et~al.} 2013, Space
  Science Reviews, 175, 125, \dodoi{10.1007/s11214-013-9960-2}

\bibitem[{Renaud \& Victoria-Feser(2010)}]{Renaud2010}
Renaud, O., \& Victoria-Feser, M.-P. 2010, Journal of Statistical Planning and
  Inference, 140, 1852, \dodoi{10.1016/j.jspi.2010.01.008}

\bibitem[{Richardson \& Cane(2010)}]{Richardson2010}
Richardson, I.~G., \& Cane, H.~V. 2010, Solar Physics, 264, 189,
  \dodoi{10.1007/s11207-010-9568-6}

\bibitem[{Rivera {et~al.}(2021)Rivera, Lepri, Raymond, Reeves, Stevens, \&
  Zhao}]{Rivera2021}
Rivera, Y.~J., Lepri, S.~T., Raymond, J.~C., {et~al.} 2021, The Astrophysical
  Journal, 921, 93, \dodoi{10.3847/1538-4357/ac1676}

\bibitem[{Rivera {et~al.}(2022{\natexlab{a}})Rivera, Raymond, Landi, Lepri,
  Reeves, Stevens, \& Alterman}]{Rivera2022}
Rivera, Y.~J., Raymond, J.~C., Landi, E., {et~al.} 2022{\natexlab{a}}, The
  Astrophysical Journal, 936, 83, \dodoi{10.3847/1538-4357/ac8873}

\bibitem[{Rivera {et~al.}(2022{\natexlab{b}})Rivera, Higginson, Lepri, Viall,
  Alterman, Landi, Spitzer, Raines, Cranmer, Laming, Mason, Wallace, Raymond,
  Lynch, Gilly, Chen, \& Dewey}]{Rivera2022a}
Rivera, Y.~J., Higginson, A., Lepri, S.~T., {et~al.} 2022{\natexlab{b}},
  Frontiers in Astronomy and Space Sciences, 9, 1056347,
  \dodoi{10.3389/fspas.2022.1056347}

\bibitem[{Schwenn(2006)}]{Schwenn2006}
Schwenn, R. 2006, Space Science Reviews, 124, 51,
  \dodoi{10.1007/s11214-006-9099-5}

\bibitem[{Shearer {et~al.}(2014)Shearer, von Steiger, Raines, Lepri, Thomas,
  Gilbert, Landi, \& Zurbuchen}]{ACE:SWICS:MLE}
Shearer, P., von Steiger, R., Raines, J.~M., {et~al.} 2014, The Astrophysical
  Journal, 789, 60, \dodoi{10.1088/0004-637X/789/1/60}

\bibitem[{Song {et~al.}(2020)Song, Zhang, Cheng, Li, Hu, Li, Chen, Zheng, \&
  Chen}]{Song2020}
Song, H.~Q., Zhang, J., Cheng, X., {et~al.} 2020, The Astrophysical Journal,
  901, L21, \dodoi{10.3847/2041-8213/abb6ec}

\bibitem[{Stakhiv {et~al.}(2015)Stakhiv, Landi, Lepri, Oran, \&
  Zurbuchen}]{Stakhiv2015}
Stakhiv, M.~O., Landi, E., Lepri, S.~T., Oran, R., \& Zurbuchen, T.~H. 2015,
  The Astrophysical Journal, 801, 100, \dodoi{10.1088/0004-637X/801/2/100}

\bibitem[{Stakhiv {et~al.}(2016)Stakhiv, Lepri, Landi, Tracy, \&
  Zurbuchen}]{Stakhiv2016}
Stakhiv, M.~O., Lepri, S.~T., Landi, E., Tracy, P.~J., \& Zurbuchen, T.~H.
  2016, The Astrophysical Journal, 829, 117,
  \dodoi{10.3847/0004-637X/829/2/117}

\bibitem[{Stone {et~al.}(1998)Stone, Frandsen, Mewaldt, Christian, Margolies,
  Ormes, \& Snow}]{ACE}
Stone, E.~C., Frandsen, A.~M., Mewaldt, R.~A., {et~al.} 1998, Space Science
  Reviews, 86, 1, \dodoi{10.1023/A:1005082526237}

\bibitem[{Subramanian {et~al.}(2010)Subramanian, Madjarska, \&
  Doyle}]{Subramanian2010}
Subramanian, S., Madjarska, M.~S., \& Doyle, J.~G. 2010, Astronomy and
  Astrophysics, 516, A50, \dodoi{10.1051/0004-6361/200913624}

\bibitem[{Tlatov {et~al.}(2014)Tlatov, Tavastsherna, \& Vasil'eva}]{Tlatov2014}
Tlatov, A., Tavastsherna, K., \& Vasil'eva, V. 2014, Solar Physics, 289, 1349,
  \dodoi{10.1007/s11207-013-0387-4}

\bibitem[{Tracy {et~al.}(2016)Tracy, Kasper, Raines, Shearer, Gilbert, \&
  Zurbuchen}]{Tracy2016}
Tracy, P.~J., Kasper, J.~C., Raines, J.~M., {et~al.} 2016, Physical Review
  Letters, 255101, 255101, \dodoi{10.1103/PhysRevLett.116.255101}

\bibitem[{Tu \& Marsch(1995)}]{Tu1995}
Tu, C.~Y., \& Marsch, E. 1995, Space Science Reviews, 73, 1,
  \dodoi{10.1007/BF00748891}

\bibitem[{Viall \& Borovsky(2020)}]{Viall:9Q}
Viall, N.~M., \& Borovsky, J. 2020, Journal of Geophysical Research: Space
  Physics, 125, 1, \dodoi{10.1029/2018JA026005}

\bibitem[{Virtanen {et~al.}(2020)Virtanen, Gommers, Oliphant, Haberland, Reddy,
  Cournapeau, Burovski, Peterson, Weckesser, Bright, van~der Walt, Brett,
  Wilson, Millman, Mayorov, Nelson, Jones, Kern, Larson, Carey, Polat, Feng,
  Moore, VanderPlas, Laxalde, Perktold, Cimrman, Henriksen, Quintero, Harris,
  Archibald, Ribeiro, Pedregosa, \& van Mulbregt}]{SciPy}
Virtanen, P., Gommers, R., Oliphant, T.~E., {et~al.} 2020, Nature Methods, 17,
  261, \dodoi{10.1038/s41592-019-0686-2}

\bibitem[{von Steiger {et~al.}(1997)von Steiger, Geiss, \&
  Gloeckler}]{vonSteiger1997}
von Steiger, R., Geiss, J., \& Gloeckler, G. 1997, in Cosmic {Winds} and the
  {Heliosphere}, Space {Science} {Series}, 581--611.
\newblock \url{https://ui.adsabs.harvard.edu/abs/1997cwh..conf..581V}

\bibitem[{Von~Steiger \& Zurbuchen(2016)}]{VonSteiger2016}
Von~Steiger, R., \& Zurbuchen, T.~H. 2016, The Astrophysical Journal, 816, 13,
  \dodoi{10.3847/0004-637X/816/1/13}

\bibitem[{von Steiger {et~al.}(2000)von Steiger, Schwadron, Fisk, Geiss,
  Gloeckler, Hefti, Wilken, Wimmer-Schweingruber, \&
  Zurbuchen}]{vonSteiger2000}
von Steiger, R., Schwadron, N.~A., Fisk, L.~A., {et~al.} 2000, Journal of
  Geophysical Research: Space Physics, 105, 27217, \dodoi{10.1029/1999JA000358}

\bibitem[{Wang \& Sheeley(2002)}]{Wang2002}
Wang, Y.~M., \& Sheeley, N.~R. 2002, Journal of Geophysical Research: Space
  Physics, 107, 1, \dodoi{10.1029/2001JA000500}

\bibitem[{Weberg(2015)}]{Weberg2015}
Weberg, M. 2015, Spatial and {Temporal} {Coordinate} {Systems} in {Space}
  {Physics}, Tech. rep.

\bibitem[{Weberg {et~al.}(2012)Weberg, Zurbuchen, \& Lepri}]{Weberg2012}
Weberg, M., Zurbuchen, T.~H., \& Lepri, S.~T. 2012, Astrophysical Journal, 760,
  \dodoi{10.1088/0004-637X/760/1/30}

\bibitem[{Widing \& Feldman(2001)}]{Widing2001}
Widing, K.~G., \& Feldman, U. 2001, The Astrophysical Journal, 555, 426,
  \dodoi{10.1086/321482}

\bibitem[{Wurz {et~al.}(2000)Wurz, Bochsler, \& Lee}]{Wurz2000}
Wurz, P., Bochsler, P., \& Lee, M.~A. 2000, Journal of Geophysical Research:
  Space Physics, 105, 27239, \dodoi{10.1029/2000JA900120}

\bibitem[{Xu \& Borovsky(2015)}]{Xu2014}
Xu, F., \& Borovsky, J. 2015, Journal of Geophysical Research: Space Physics,
  120, 70, \dodoi{10.1002/2014JA020412}

\bibitem[{{Yogesh} {et~al.}(2021){Yogesh}, Chakrabarty, \&
  Srivastava}]{Yogesh2021}
{Yogesh}, Chakrabarty, D., \& Srivastava, N. 2021, Monthly Notices of the Royal
  Astronomical Society: Letters, 503, L17, \dodoi{10.1093/mnrasl/slab016}

\bibitem[{{Yogesh} {et~al.}(2023){Yogesh}, Chakrabarty, \&
  Srivastava}]{Yogesh2023}
---. 2023, Monthly Notices of the Royal Astronomical Society, 526, L13,
  \dodoi{10.1093/mnrasl/slad112}

\bibitem[{Zerbo \& Richardson(2015)}]{Zerbo2015}
Zerbo, J.-L., \& Richardson, J.~D. 2015, Journal of Geophysical Research: Space
  Physics, 120, 10,250, \dodoi{10.1002/2015JA021407}

\bibitem[{Zhao {et~al.}(2022)Zhao, Landi, Lepri, \& Carpenter}]{Zhao2022}
Zhao, L., Landi, E., Lepri, S.~T., \& Carpenter, D. 2022, Universe, 8, 393,
  \dodoi{10.3390/universe8080393}

\bibitem[{Zhao {et~al.}(2017{\natexlab{a}})Zhao, Landi, Lepri, Gilbert,
  Zurbuchen, Fisk, \& Raines}]{Zhao:InSituComposition:Sources}
Zhao, L., Landi, E., Lepri, S.~T., {et~al.} 2017{\natexlab{a}}, The
  Astrophysical Journal, 846, 135, \dodoi{10.3847/1538-4357/aa850c}

\bibitem[{Zhao {et~al.}(2017{\natexlab{b}})Zhao, Zhang, \& Rassoul}]{Zhao2017a}
Zhao, L., Zhang, M., \& Rassoul, H.~K. 2017{\natexlab{b}}, The Astrophysical
  Journal, 836, 31, \dodoi{10.3847/1538-4357/836/1/31}

\bibitem[{Zurbuchen \& Richardson(2006)}]{Zurbuchen2006}
Zurbuchen, T.~H., \& Richardson, I.~G. 2006, Space Science Reviews, 123, 31,
  \dodoi{10.1007/s11214-006-9010-4}

\bibitem[{Zurbuchen {et~al.}(2016)Zurbuchen, Weberg, Von~Steiger, Mewaldt,
  Lepri, \& Antiochos}]{Zurbuchen2016}
Zurbuchen, T.~H., Weberg, M., Von~Steiger, R., {et~al.} 2016, The Astrophysical
  Journal, 826, 10, \dodoi{10.3847/0004-637X/826/1/10}

\end{thebibliography}
\bibliographystyle{aa}

\end{document}